\documentclass[aps,pre,floatfix,superscriptaddress,twocolumn,10pt]{revtex4-2}
\usepackage{color}
\usepackage{graphicx}
\usepackage{amsmath, amsthm, amssymb}
\usepackage[colorlinks,citecolor=red,urlcolor=blue]{hyperref}
\newcommand{\be}{\begin{equation}}
\newcommand{\ee}{\end{equation}}
\newcommand{\bea}{\begin{eqnarray}}
\newcommand{\eea}{\end{eqnarray}}

\begin{document}
\title{Entropy of fully packed hard rigid rods on $d$-dimensional hyper-cubic  lattices}

\author{Deepak Dhar}
\email{deepak@iiserpune.ac.in}
\affiliation{Indian Institute of Science Education and Research, Dr. Homi Bhabha Road, Pashan,  Pune 411008, India}
\author{R. Rajesh} 
\email{rrajesh@imsc.res.in}
\affiliation{The Institute of Mathematical Sciences, C.I.T. Campus, Taramani, Chennai 600113, India}
\affiliation{Homi Bhabha National Institute, Training School Complex, Anushakti Nagar, Mumbai 400094, India}

\date{\today}

\begin{abstract}
We determine the asymptotic behavior of the entropy of  full coverings of a $L \times M$ square lattice by rods of size $k\times 1$ and $1\times k$, in the limit of large $k$.
We show that full coverage is possible only if at least one of $L$ and $M$ is a multiple of $k$, and that all allowed configurations can be reached from a standard configuration of all rods being parallel, using only basic flip moves that replace a $k \times k$ square of parallel horizontal rods by vertical rods, and vice versa. In the limit of  large $k$, we show that the entropy per site $S_2(k)$ tends to $ A k^{-2} \ln k$, with $A=1$. We  conjecture, based on a perturbative series expansion, that this large-$k$ behavior of entropy per site is super-universal and continues to hold on all $d$-dimensional hyper-cubic lattices, with $d \geq 2$.

\end{abstract}

\maketitle

\section{Introduction}

Systems of particles with only hard core interactions between them have been studied as prototypical models for phase transitions in equilibrium statistical mechanics as well as for understanding aspects of non-equilibrium statistical mechanics.  
In equilibrium statistical mechanics, hard sphere systems serve as minimal models of solid to fluid transition in molecular solids~\cite{1957-aw-jcp-phase,1962-aw-pr-phase,noya2008determination}, and in colloidal crystals~\cite{1986-pm-nature-phase}. 
Dimer models are equivalent  to the Ising model, and 
anisotropic  hard particles can effectively model different phases and phase transitions in liquid crystals~\cite{fisher1966dimer,2003-hkms-prl-coulomb,mossner2003ising,1995-oup-gp-physics,frenkel1988structure,care2005computer}. In non-equilibrium statistical mechanics, hard core models like symmetric or asymmetric exclusion processes provide basic models for driven systems and jamming in granular
systems~\cite{donev2004jamming,liggett2012interacting,mallick2015exclusion}.

Lattice models of hard-core  particles have been of particular interest, as they are analytically more tractable. The phases of assemblies of particles of many different shapes have been studied. Examples include squares~\cite{1966-bn-prl-phase,1967-bn-jcp-phase,2012-rd-pre-high,2015-rdd-prl-columnar,2016-ndr-epl-stability,mandal2017estimating}, triangles~\cite{1999-vn-prl-triangular}, hexagons~\cite{1980-b-jpa-exact}, long rods~\cite{1956-f-prs-phase,2007-gd-epl-on,2011-drs-pre-hard}, rectangles~\cite{2014-kr-pre-phase,2015-kr-epjb-phase,2015-kr-pre-asymptotic}, Y-shaped molecules~\cite{szabelski2013self,2015-rthrg-tsf-impact,2018-pre-mnr-phase}, tetraminoes~\cite{2009-bsg-langmuir-structure}, lattice gases with  exclusion upto $k$th nearest neighbors~\cite{2005-p-jcp-thermodynamic,2007-fal-jcp-monte,2014-nr-pre-multiple,2016-nr-jsm-high,akimenko2019tensor,PhysRevE.101.062138}, cubes~\cite{vigneshwar2019phase}, plates~\cite{disertori2020plate}, etc. An analytical exact solution has been possible only for the case of hard hexagons so far~\cite{1980-b-jpa-exact}.  Phase transitions have also been studied in mixtures of different shapes, for example squares and dimers~\cite{2015-rdd-prl-columnar,mandal2017columnar},  rods of different lengths~\cite{2015-ksr-epl-phase,2015-sr-pre-polydispersed}, polydispersed spheres~\cite{rodrigues2019three}, etc.  For the mixture of squares and dimers, it was shown that the critical exponents of the order-disorder transition depends continuously on the relative concentration of the components. Despite a long history, many basic questions about these systems remain open; for example, for a given shape of particles, what are the possible ordered phases, and in which sequence will they appear on increasing the density?

System of hard rods/cylinders have attracted a lot of interest,
starting with the pioneering work of Onsager, who showed that a system of thin, long cylinders in three dimensional continuum undergo a phase transition from a disordered phase to an orientationally ordered nematic phase~\cite{1949-o-nyas-effects}. The study of lattice models of linear $k\times 1$ hard rods ($k$-mers) started with the work of  Flory~\cite{1956-f-prs-phase} and Zwanzig~\cite{1963-z-jcp-first}.
On a $d$-dimensional hyper-cubic lattice, rods can only orient  in one of the $d$ directions. It was realized in Ref.~\cite{2007-gd-epl-on}, based on Monte Carlo simulations and high density expansions, that nematic order is present at intermediate densities for large enough $k$, and that the lattice model at high densities must undergo a second disordering transition at a critical density $1-\rho_c \sim k^{-2}$ for large $k$, when the nematic order is lost. Usual  Monte Carlo  techniques with local moves are rather inefficient in sampling states at  high-density  due to high rates of rejection of moves due to jamming, but recently-introduced strip-update Monte Carlo technique has made it possible to reach densities within a few percent of maximum packing density~\cite{2012-krds-aipcp-monte,2013-krds-pre-nematic}.  Using these techniques, it is found that on the square lattice, for $k < 6$, there is no phase transition, but for $k>6$, as density is increased, there are three phases: the low-density disordered phase, intermediate-density nematic phase, and the high-density phase in which there is no long ranged positional or orientational order~\cite{2013-krds-pre-nematic}. The existence of the transition may be rigorously proved~\cite{2013-dg-cmp-nematic}. The first phase transition belongs to the Ising~\cite{2008-mlr-epl-determination,2008-mlr-jcp-critical,2009-fv-epl-restricted} or three-state Potts universality classes~\cite{2008-mlr-epl-determination,2008-mlr-jcp-critical,2008-mlr-pa-critical} depending on whether the rods are on a square or triangular/honeycomb lattice. The nature of the second transition is not so clear. There is some indication of the high density phase having power law correlations~\cite{2013-krds-pre-nematic} with the second transition not being in the Ising universality class~\cite{2013-krds-pre-nematic,chatelain2020absence}, while the exact solution of  soft repulsive rods on a tree-like lattice~\cite{2013-kr-pre-reentrant} suggests otherwise. More recently, the transitions in two dimensions have been studied using measures such as  the  classical  `entanglement' entropy,  mutability, Shannon entropy and data compression~\cite{vogel2017phase,chatelain2020absence,vogel2020alternative}. 

In three dimensions, there is no phase transition for $k \leq 4$. For $k\geq 7$, the system undergoes phase transitions from disordered to nematic to a layered disordered phase as density is increased. In the layered disordered phase, the system breaks up into very weakly interacting two dimensional planes within which the rods are disordered. For $4 < k< 7$, there is no nematic phase, and a single phase transition from a disordered to a layered disordered phase~\cite{2017-vdr-jsm-different,2017-gkao-pre-isotropic}. 

In this paper, we focus on the fully packed limit of linear rods on the square lattice and give heuristic arguments to extend the results to higher dimensions.  In particular, we focus on the entropy per site $S_d(k)$. We note that, in addition to  understanding phase transitions in lattice systems of rods with finite density of vacancies, the study of the fully packed phase is relevant for other physical systems. For instance, it would help understand the tetratic order in self assembly of squares and rectangles in the continuum~\cite{2004-wf-cmst-tetratic}.  It  may also help in our understanding of phase transitions in other strongly correlated systems.  For example, the binding unbinding transition  of quarks  as a function of the density of hadrons in nuclear matter,  studied in QCD, is similar to  the binding-unbinding transition of $k$ species of holes during the transition from the disordered high density phase to the nematic  phase of $k$-mers.

For the case of dimers ($k=2)$ - the only case that is exactly solvable~\cite{1961-k-physica-statistics,kasteleyn1963dimer,1961-tf-pm-dimer,1961-f-pr-statistical} - the entropy per site for square lattice is $S_2(2)=G/\pi=0.29156\ldots$, where $G$ is the Catalan's constant~\cite{1961-k-physica-statistics}. On square lattice, 
the orientation-orientation correlation of two dimers separated by a distance $r$ decays as a power law $r^{-1/2}$ for large $r$~\cite{1963-fs-pr-statistical}, while on a triangular lattices, these correlations are short-ranged~\cite{2002-fms-prb-classical}. A review of the method of solution of dimer problems on planar lattices may be found in Ref.~\cite{nagle1989dimer}. In three dimensions, there is a class of lattices (not cubic lattice) for which an exact solution can be found and the correlations are strictly finite ranged~\cite{2008-dc-prl-exact}, while for the cubic lattice, the orientational correlations decay as a power law~\cite{2003-hkms-prl-coulomb}. The entropy of fully packed trimer ($k=3$) tilings on square lattice have also been studied~\cite{2007-gdj-pre-random}. By numerically diagonalising  the transfer matrices for strips, the entropy per site was found to be $S_2(3)=0.158520 \pm 0.000015$. Much less is known for higher values of $k$. It is known that the tilings admit a vector height field representation~\cite{kenyon2000conformal}. 
For larger values of $k$, Gagunashvili and Priezzhev obtained an upper bound for the entropy on the square lattice: $S_2(k) \leq k^{-2} \ln(\gamma k)$, where $\gamma= \exp(4 G/\pi)/2$, with $G$ being the Catalan's constant~\cite{gagunashvili1979close}.
It is clear that the full-packing constraint induces strong correlations in the orientations of rods, and one would generally expect orientation-orientation correlations to decrease with distance  as a power-law.  

The full packing constraint
severely limits the  allowed configurations.  One way to generate a large number of such configurations, satisfying  all these constraints is to break the system into parallel 2-dimensional layers, and fully pack each layer with rods. Since rods on different layers do not interact, configurations on different layers can be independently   generated, giving a large entropy.   Indeed, there is evidence 
from Monte Carlo simulations (the simulations are done not at fully packing, but for densities close to full-packing) that the  high density phase of long rods in three dimensions shows two-dimensional layering~\cite{2017-vdr-jsm-different}, and  our perturbation expansion suggests that,  in the fully packed limit, configurations in even higher dimensions would be dominated by layered two-dimensional configurations.

In this paper, we determine the asymptotic behavior of the entropy of the fully packed configurations in the limit of large rod lengths $k$: first  in two dimensions, and then generalized to higher dimensions. The number of coverings depends strongly on the boundary conditions imposed. 
We will consider configurations of a finite  $L \times M$  rectangular portion of square lattice fully covered by rectangles of size $k \times 1$ or $1 \times k$. Equivalently, we can consider this  a lattice model, with all sites  covered using  straight rigid rods of length $k$. We will call this  open boundary conditions.   We prove  that full coverage  in the open boundary case is possible only if at least one of $L$ and $M$ is a multiple of $k$.  All the allowed configurations for this case can be reached from the standard configuration of all horizontal rods, using only basic flip moves that flip a $k \times k$ square of parallel horizontal rods by vertical rods, and vice versa.  
Using rigorous upper and lower bound estimates, we show that $S_2(k)$, to leading order in $k$, equals  $A k^{-2} \ln k$ with $A=1$.   

Based on a perturbation series expansion, we conjecture that in higher dimensions, the entropy for the fully packed phase, for large $k$, would be dominated by configurations where the rods arrange themselves in stacked two dimensional layers.  Thus, we conjecture that the large-$k$ behavior of entropy per site is `super-universal', and continues to hold on $d$-dimensional hypercubical lattices for all $d > 2$ and 
\begin{equation}
\lim_{k \rightarrow \infty} \frac { k^2 S_d(k)}{ \ln k} = 1,
\label{eq:sd}
\end{equation}
independent of $d$.

The remainder of the paper is organized as follows.  In Sec.~\ref{sec:prelims}, we define the problem precisely. We derive some basic properties of the fully packed phase by showing that an $L \times M$ rectangle can be completely covered by $k$-mers, only if at least one of $L$ or $M$ is multiple of $k$ and that all full packing configurations on an open $L \times M$ rectangle can be obtained from the standard configuration of all horizontal rods by a combination of basic flip moves.  In Sec.~\ref{sec:lowerbound}, we obtain lower bounds for entropy by solving exactly for the entropy of rods on semi-infinite strips   $k \times \infty$ and $2 k \times \infty$. These results are generalised to arbitrary strips $lk \times \infty$ by considering truncated generating functions. In Sec.~\ref{sec:upperbound}, we combine the lower bounds for entropy with existing upper bounds  to obtain Eq.~(\ref{eq:sd}).  In Sec.~\ref{sec:higherdim}, we use heuristic arguments based on perturbation theory to support the conjecture that  that this result should also hold for  all $d$-dimensional hypercubical lattices with $d >2$. Section~\ref{sec:conclusions} contain some concluding remarks.

\section{\label{sec:prelims} Preliminaries}

\begin{figure}
\includegraphics[width=0.8\columnwidth]{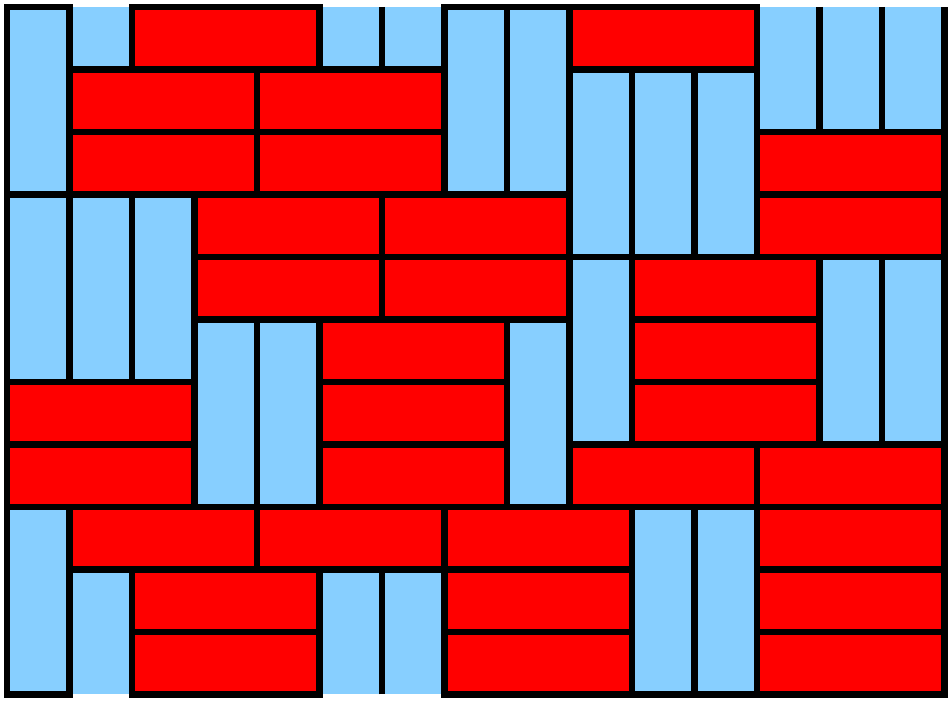}
\caption{\label{fig:tilingexample} A tiling of the Euclidean plane by $k$-mers, with $k=3$.  Only a part of the tiling is shown here, and some $k$-mers do not fully fall in the  region shown.  }
\end{figure}

We consider  tilings of a $L \times M$ rectangle, with $L$, $M$ positive  integers, by  $k \times 1$ and $1\times k$ rectangles ($k$-mers).  Each $k$-mer can only be in one of two orientations:  horizontal or vertical. An example is shown in Fig.~\ref{fig:tilingexample} for the case $k=3$. Equivalently, we can consider this  a lattice model, with all sites  covered using  straight rigid rods of length $k$. Let $N(L,M)$ be the number of such tilings.  

\subsection{\label{subsection:divisibility}Divisibility of $L$, $M$ by $k$}

We first show that $N(L,M)$ is non-zero, if and only if at least one of $L$ and $M$ is  divisible by $k$.   The `if' part is trivial. For the other part, clearly $LM$ has to be a multiple of $k$, for full coverage. We now argue that in this case, at least one of $L$ and $M$ has to be a multiple of $k$.

\begin{figure}
\begin{center}
\includegraphics[width=0.8\columnwidth]{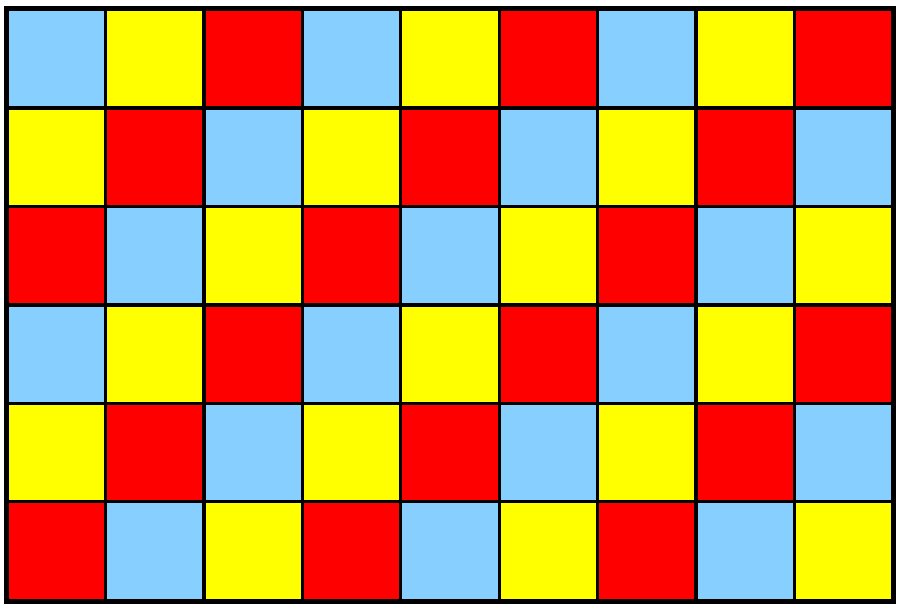}
\end{center}
\caption{\label{fig:colors} Assigning colors to $1\times 1$ squares for the case $k=3$.}
\end{figure}

\begin{figure}
\begin{center}
\includegraphics[width=\columnwidth]{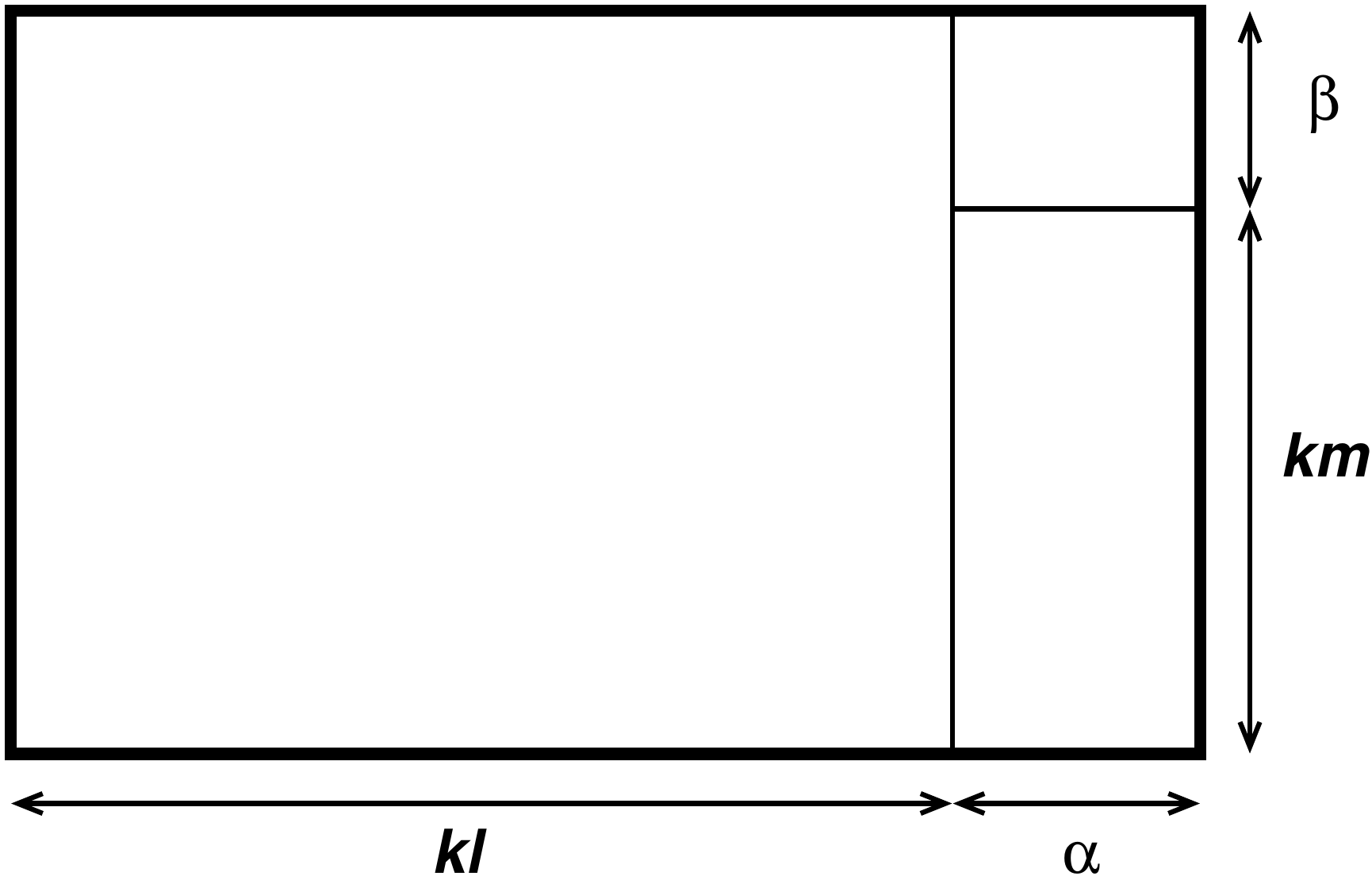}
\caption{\label{fig.subrectangles} Dividing an $(k \ell +\alpha) \times(k m +\beta)$ rectangle, where $0 < \alpha, \beta \leq k-1$, into smaller rectangles.}
\end{center}
\end{figure}

Assign one of the $k$ colors, called here $0,1,2, \ldots, (k-1)$ to each of the squares of the lattice, with square $(x,y)$ given color $ q = (x -y) \mod k$. The coloring of the squares for the case $k=3$ is shown in Fig.~\ref{fig:colors}. Then each $k$-mer covers exactly one square of each color. Let $L = k ~\ell + \alpha$,  $M = k~m + \beta$, with $0 < \alpha, \beta \leq k-1$. 
Divide the rectangle into three smaller rectangles of sizes $k \ell \times M, \alpha \times k m$ and $\alpha \times \beta$, as shown in Fig.~\ref{fig.subrectangles}. Then, clearly the rectangles of size $k \ell \times M$, and $\alpha \times k m$ can be covered by $k$-mers, implying that the number of squares of different colors in these two rectangles are equal.  However, the small rectangle of size $\alpha \times \beta$ has $\min(\alpha, \beta)$ squares of same color along the diagonal.  To cover them would require at least $\min(\alpha, \beta)$ rods, with total area  $k \min(\alpha, \beta)$. Equating this to the total area $\alpha \beta$, we obtain $k=\max(\alpha, \beta)$. This contradicts the assumption that $\alpha, \beta  < k$. Hence, the rectangle can not be fully covered by $k$-mers, unless either $L$ or $M$ is divisible by $k$.

For simplicity of presentation, in the following, we shall  assume that both $L$ and $M$ are multiples of $k$.

\subsection{\label{subsection:ergodicity} Ergodicity of the flip moves}

In this subsection, we show that  all configurations of rods can be reached  from any configurations by just using the flip move (defined below).

\begin{figure}
\centering
\includegraphics[width=\columnwidth]{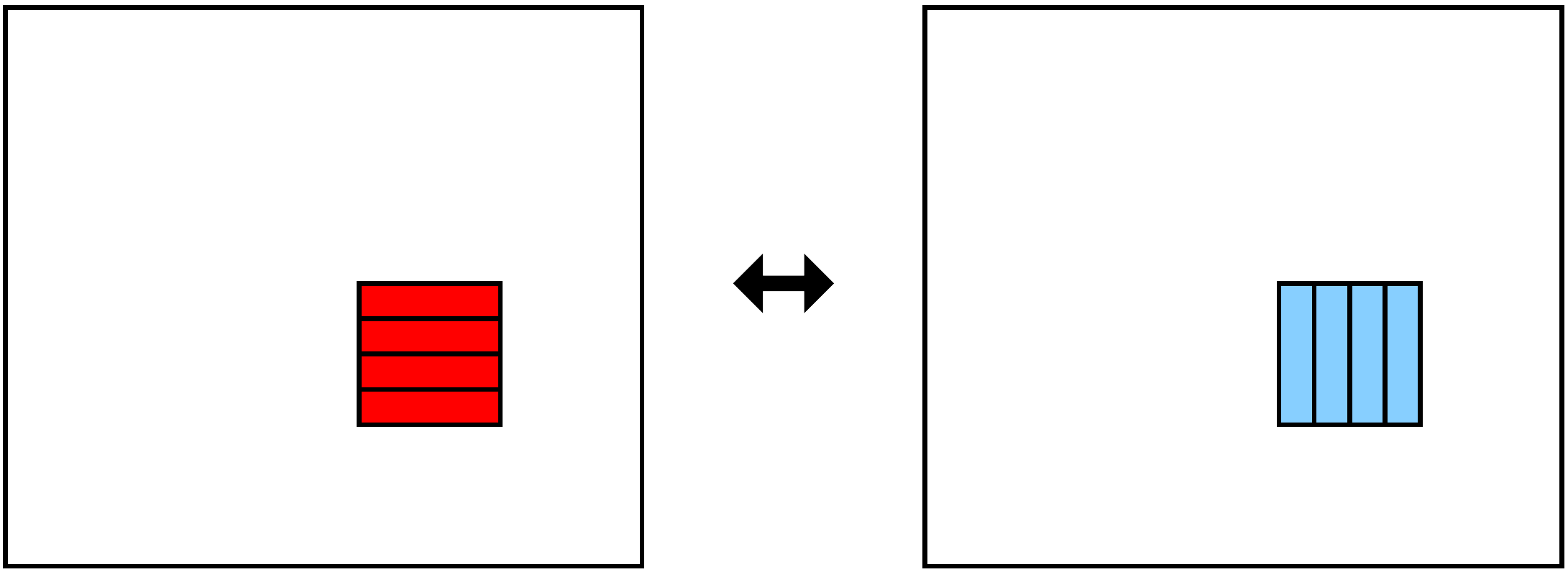}
\caption{\label{fig:flipmove}The basic flip moves consists of replacing a small $ k \times k$ square in the configuration covered  by $k$ horizontal $k$-mers, by vertical $k$-mers, and vice versa.}
\end{figure}

We define the standard tiling configuration of $k \ell \times M$ rectangle by $k$-mers as one using only horizontal $k$-mers. 
A basic flip move is defined as replacing a  $k \times k$  square filled with vertical $k$-mers by one with horizontal $k$-mers, and vice versa, as illustrated in Fig.~\ref{fig:flipmove}.

\begin{figure}
\includegraphics[width=\columnwidth]{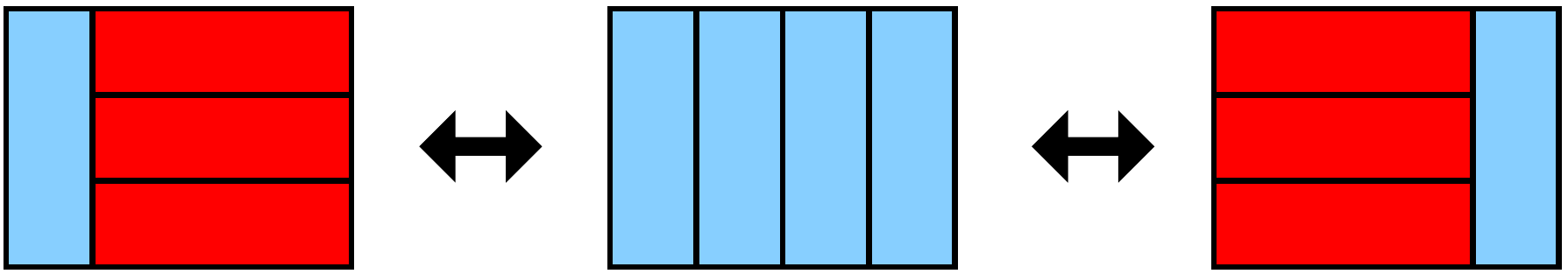}
 \caption{\label{fig:slidemove}The slide move consists of transposing a rod and an adjacent  flippable square, i.e. sliding the rod across the square, may be thought of  as a combination of two flip moves. }
\end{figure}

A combination of  two flip moves defines a `slide' move, where a vertical $k$-mer next to a $k \times k$ flippable square exchanges position (see Fig.~\ref{fig:slidemove}), and the vertical $k$-mer will be said to slide across the flippable square.  

We now argue that that any full tiling of $k \ell \times M$ rectangle by $k$-mers may be reached from the standard configuration by using only the basic flip  and slide moves.

\begin{figure}
\begin{center} 
\includegraphics[width=\columnwidth]{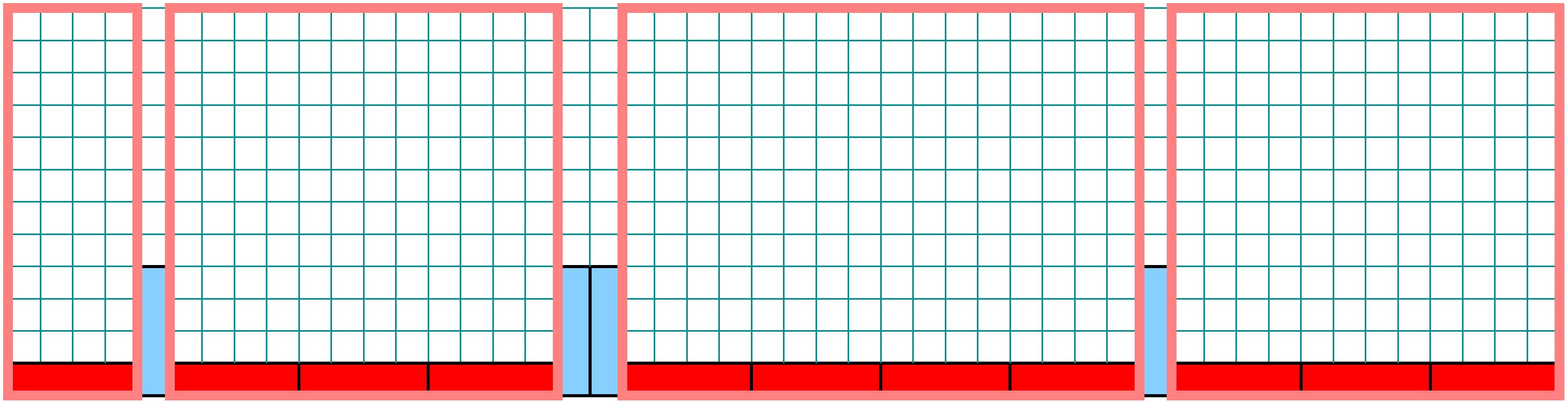}
\caption{\label{fig:fliipableblocks}A tiling of a $48\times 12$ rectangle with  rods of length $k=4$, where only the rods in the bottom row are shown. The vertical rods split the rectangles into smaller rectangles, and aids in finding a block of flippable $k$-mers (see text for details).} 
\end{center} 
\end{figure}

{\bf Proof:} Look at the lowest row. If it consists of only horizontal $k$-mers, then we  ignore this row, and the problem reduces to one with a smaller $M$.
Else, it would have $\ell' = \ell -\Delta$ horizontal $k$-mers, and $k \Delta$ vertical $k$-mers.  In Fig.~\ref{fig:fliipableblocks}, we have shown an example of a $4$-mer tiling of a  $48 \times 12$ rectangle, where  $\ell =12$, $\Delta =1$. 
We move to the left  any $k \times k$  block of horizontal flippable rods we find  between these $k \Delta$ vertical $k$-mers,  using the slide move, and make the vertical rods closer to each other.
If  now there is any block of consecutive vertical $k$-mers, we can flip these to horizontal, and reduce the problem to one with fewer number of vertical 
$k$-mers.  

If there is no such horizontal flippable block of rods, we look at the bottom row. Let us say that it has segments of $i_1, i_2, ... i_s$ horizontal rods, interspersed with vertical rods.  [In Fig.~\ref{fig:fliipableblocks}, there are $4$
segments, with  $i_1=1$, $i_2=3$, $i_3 =4$, $i_4 =3$]. Clearly, these are bordered by vertical $k$-mers at the ends, unless the segment itself is at the end of the rectangle.  Then we look at the sub-rectangles of sizes $i_1 k \times M, i_2 k\times M,..$ made up of these segments and bounded by vertical boundaries. In the example shown in Fig.~\ref{fig:fliipableblocks}, these rectangles are shown with orange boundaries.   

We now argue that there will be a flippable $k \times k$ block within each of these small rectangles.    This is clear if the width of the rectangle is exactly $k$. Then the sites just above can only be covered by a horizontal rod, or $k$ vertical rods. In the latter case, it forms a vertical flippable rectangle. If not,  then eventually, we will have $k$ horizontal rods just above each other,   and form a horizontal flippable rectangle. 

If the width is greater than $k$, and the row just above is not made of all horizontal rods, then it will be made up of a number of horizontal segments, separated by vertical rods. And we can repeat the argument with this smaller set.  This process can not continue for ever, as the total width is finite, and the width decreases at each step.  

Thus, we will be able to find a flippable $k \times k$ box at each stage, and eventually, the number of vertical rods becomes zero, and the standard tiling of all horizontal $k$-mers is reached. Since all moves are reversible, and  any valid configuration of full-packing   can be changed to standard configuration, we can go from any full packing configuration on the rectangle to any other using only flip moves.

\section{\label{sec:lowerbound} Lower bound for entropy for large $k$}

\begin{figure}
\begin{center}
\includegraphics[width=0.6\columnwidth]{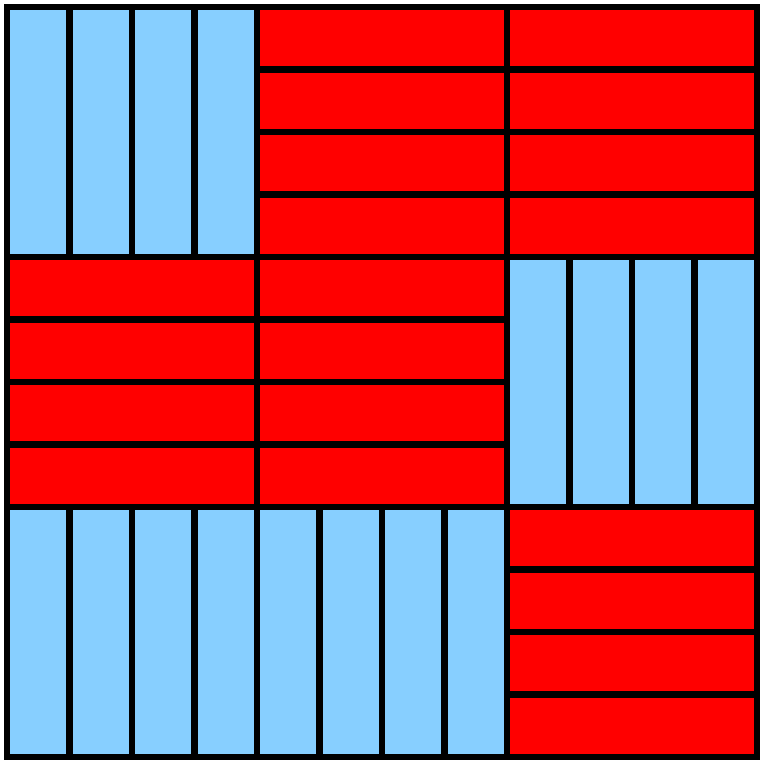}
\caption{\label{fig:dividing} Dividing a rectangle into  $k \times k$ squares (here, $k=4$). }
\end{center}
\end{figure}

We first show  that at full packing, there is a finite entropy per site.  We divide the lattice
into $k \times k$ squares. There are $LM/k^2$ such squares, and each can be tiled in two ways, independent of the others. Then the total number of such  tilings is $2^{LM/k^2}$ (see Fig.~\ref{fig:dividing}). Of course, more complicated tilings are possible, as shown in Fig.~\ref{fig:tilingexample}, and the above only provides a lower bound.  
We define entropy per site 
\begin{equation}
S_2(k) = \lim_{L, M \rightarrow \infty} \frac{\ln N( L, M)}{ LM}.
\end{equation} 
Then, $S_2(k)  \geq k^{-2} \ln 2$.

\subsection{\label{subsection:kxinfinity} Entropy of strips $k \times \infty$}

We can easily  obtain a better lower bound on $S_2(k)$.  Break the $L
\times k M$ lattice in $M$ strips of width $k$ each. Let $N(L,k)$ be denoted by $F_L$.  Since by breaking into strips of width $k$, we disallow configurations where rods cross the boundary, leading to undercounting, we obtain the inequality
\begin{equation}
 N(L,kM) \geq [ F_{L}]^M. 
\end{equation}
$F_L$ obeys a simple recursion relation. Consider the packing of a  $k \times L$ rectangle. The first row can be covered by a horizontal $k$-mer (reducing $L$ by one) or the first $k \times k$ square can be covered $k$ parallel vertical $k$-mers  (reducing $L$ by $k$). Thus, 
$F_L$'s satisfy the recursion relation
\begin{equation}
F_L = F_{L-1} + F_{L-k}. \label{eq1}
\end{equation}
This implies that $F_L$ increases as $\lambda^L$ where $\lambda$ is
the largest root of the equation
\begin{equation}
\lambda^k = \lambda^{k-1} + 1. \label{eq2}
\end{equation}
For large $k$, to leading order $\lambda =1$.  A little bit of algebra shows that
In the limit of  large $k$, the subleading terms take the form
\begin{equation}
\lambda = 1 + \frac{W(k)}{k} + \mathcal{O} \left( k^{-2} \right),
\end{equation}
where $W(k)$ is the solution of the equation 
\begin{equation}
W(k) \exp[ W(k)] = k.
\label{eq:lambert1}
\end{equation}
The function $W(k)$ is called the Lambert function~\cite{corless1996lambertw}. To leading order, $W(k) \sim \ln k$ for large $k$ (to see this, take logarithm on both sides of Eq.~(\ref{eq:lambert1}) and compare the terms of leading order). The sub-leading term can be similarly obtained to give for large $k$
\begin{equation}
W(k) \approx \ln\left(\frac{k}{\ln k}\right), 
\label{eq:widhtk}
\end{equation}
with corrections that only grow  slower than  $\ln( \ln k)$. Thus we obtain
\begin{equation}
\lambda = 1 +\frac{1}{k} \ln\left(\frac{k}{\ln k} \right) + {\rm 
higher~order~terms}. \label{eq3}
\end{equation}
The 
entropy per site for the $k\times \infty$ strip is  $(\ln \lambda) / k$. Thus,
\be
S_{k\times \infty} = \frac{\ln k}{k^2} \left(1-\frac{\ln\ln k}{\ln k} + \ldots\right)
\label{eq:10}
\ee
Since $S_{k\times \infty}$ is a lower bound for $S_2(k)$, we obtain the leading behavior:
\begin{equation}
\lim_{k\to \infty} \frac{k^2 S_2(k)}{\ln k} \geq 1. \label{eq4}
\end{equation}

\subsection{\label{subsection:2kxinfinity} Entropy of strips $2 k \times \infty$}

In this subsection, we describe the exact calculation of the entropy of tilings of the semi-infinite $2 k \times \infty$ stripe with $k$-mers, where the  $y$-coordinate  is 
$\geq 1$, and the $x$-coordinate lies in the range $[1,2k]$.  
We define the generating function $\Omega_{2k}(x)$ as the sum over all covering of rectangles of size $2k \times r$, summed over all positive integer values of $r$, where the weight of a covering with $n$ tiles is $z^n$.   Then, we have 
\begin{equation}
\Omega_{2k}(z) = \sum_{r=0}^{\infty}   N(2k, r) z^{2r}.
\end{equation}

\begin{figure} 
\begin{center}
\includegraphics[width=0.5\columnwidth]{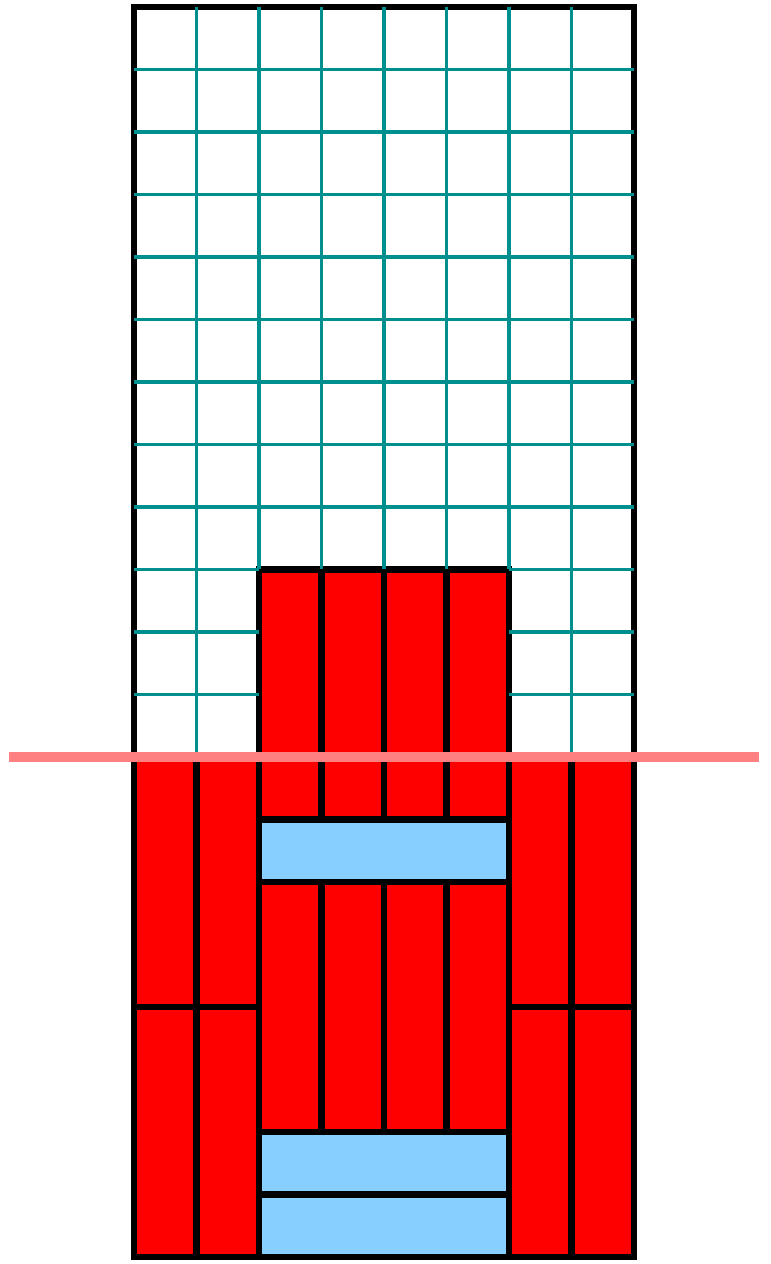}
\end{center}
\caption{\label{fig:partialfilling} A partial filling of the strip $2k \times \infty$ by rods for the case $k=4$. The  horizontal red line shows the reference line. The boundary  of the configuration is specified by the projection to the top of the reference line; in this case $\{ 0,0,3,3,3,3,0,0\}$, or in  a more compact notation, denoted as $\{ 0^2 3^4 0^2\}$.  }
\label{fig:fig1} \end{figure}

We also define a  partial covering of the strip with rods to the  bottom of some reference line $y = s >0$, 
so that no site  with $y$-coordinate less than $s$ is left uncovered (see Fig.~\ref{fig:partialfilling}), and all rods must cover at least one site with $y$-coordinate less than $s$.   Clearly, all rods that do not lie completely to the bottom of the $y=s$ must be vertical.  A partial covering is a rectangular  covering iff no site with $y$-coordinate larger than $s$ is covered. 

A  partial covering  may be characterized by its top boundary $\{h_x\}$, for $x = 1 $ to $2 k$, where $h_x$ specifies how many sites to the top of the reference line   $y=s$  are covered in the column with coordinate $x$.   We will choose $s$ to be as large as possible, so that at least one of the $h_x$'s has to be zero, and $h_x \leq k-1$, for all $x$. For example, the boundary  of the configuration shown in Fig.~\ref{fig:partialfilling} is specified by $\{ 
0,0,3,3,3,3,0,0\}$. In  a more compact notation, we will write this as $\{ 0^2 3^4 0^2\}$.

Not all height configurations are allowed. A bit of thought shows that for a partial covering of the $2 k \times L$ stripe, the only allowed height configurations are $\{ 0^{2k}\}$,  $\{ h^k 0^k\}$, $\{ 0^k h^k \}$,   
$\{ h^j 0^k h^{k-j}\}$ or  $\{ 0^j h^k 0^{k-j}\}$, with $ h $ and $j$ taking values 
from $ 1$ to $k-1$.

We 
define the 
generating functions $\psi(\{h_j\})$ as the generating function of all 
possible ways of completing a partial 
tilings with a given height profile $\{ h_j\}$, where the completed covering is rectangular, and the weight of tiling in which we add $n$ extra  rods is $z^n$.  Therefore,  for example, 
\begin{eqnarray}
\psi(\{0^{2 k}\}) &=& \Omega_{2k}(z)= 1+z^2+z^4 + \ldots,\\
\psi(\{0^k 1^k\}) &= &z + z^3 + \ldots. \label{eq5}
\end{eqnarray}

Consider a particular height configuration $\{h_x\}$. We can write 
recursion equations for the corresponding generating function 
$\psi(\{h_x\})$, by  considering all possible ways of filling the column 
of sites immediately to the top of the reference line by $k$-mers, 
such that the 
top edge of the full tiling is horizontal, and no sites are left 
uncovered. 

\begin{figure} 
\begin{center}
\includegraphics[width=\columnwidth]{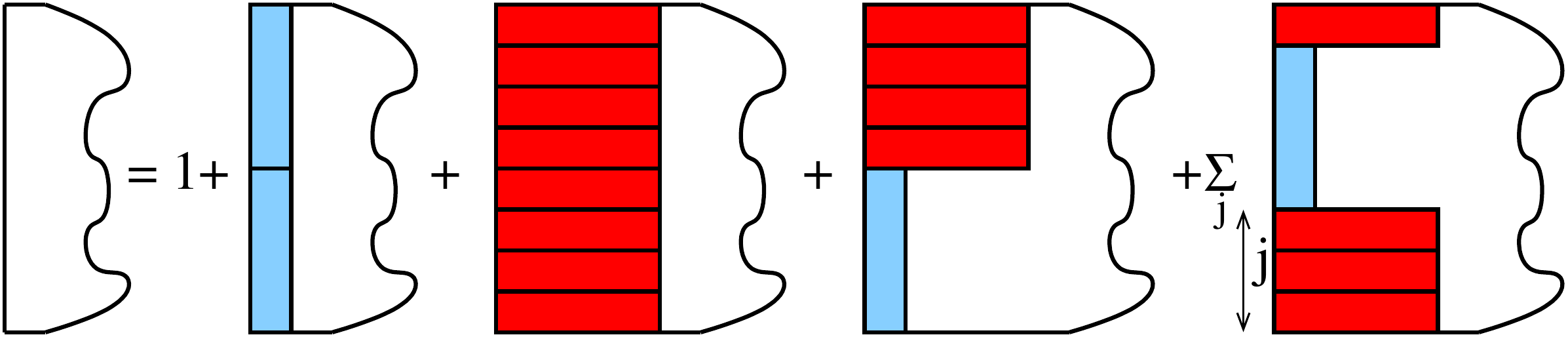}
\end{center}
\caption{\label{fig:recursion}Recursion equation for $\Psi(\{0^{2k}\})$, shown for $k=4$.  Each subfigure is rotated clockwise by 90 degrees for conserving space. The jagged boundary at the right end indicates summing over all possible configurations on the right. }
\end{figure}

For example, it is easily seen that  (see Fig.~\ref{fig:recursion}) 
\begin{eqnarray}
\Psi(\{0^{2k}\}) &=& 1 + (z^2 + z^{2k}) \Psi(\{0^{2k}\}) \nonumber\\
&&+ 2 z^{k+1} \Psi(\{0^{k}(k-1)^k\}) \label{eq:A4}\\
&&+\sum_{j=1}^{k-1} z^{k+1} \Psi(\{(k-1)^j 0^k (k-1)^{k-j}\}). \nonumber
\end{eqnarray}
The different terms in this equation correspond to the cases where the 
next row is left empty, or filled by two horizontal rods, or by $2 k$ 
vertical rods, or by first $j$ vertical rods, then a horizontal rod, 
then $k-j$ vertical rods. 

Writing such generating functions for all possible height configurations, we obtain a 
set of inhomogeneous linear equations in  approximately $2  k^2  $ 
variables. This 
may be written as a transfer matrix of dimension $2 k^2  \times 2 k^2$.   
However,  using the symmetries of the problem, this number 
can be considerably reduced.

We note that the recursion equations for    the generating function $\Psi(\{h^j 0^k h^{k-j}\}$ 
is   
\bea
&&\Psi(\{h^j 0^k h^{k-j}\} =  z \Psi(\{(h-1)^j0^k (h-1)^{k-j}\}) \nonumber \\
&& + z^k \Psi(\{0^j (k-h)^k 0^{k-j}\}), ~~
1 \leq h < k. 
\label{eq:A5}
\eea
This equation has no 
$j$-dependence. Hence, we may expect that $\Psi(\{h^j 0^k h^{k-j}\}$ is 
independent of $j$.  It can be checked that this ansatz is consistent with 
the remaining recursion equations.   Similarly, we find that $\Psi(\{0^j 
h^k 0^{k-j}\})$ is also independent of $j$.  With this simplification, 
the number of independent variables reduces to approximately $2 k $.  

The remaining recursion equations are easily written down, we obtain for all $ 1 \leq h \leq k-1$,
\begin{equation} \Psi(\{0^j h^k 0^{k-j}\}) = z^k \Psi(\{(k-h)^j 0^k 
(k-h)^{k-j}\}), \label{eq:A6} \end{equation} 
and 
\begin{equation} 
\Psi(\{0^k h^k\}) = z \Psi(\{0^k (h-1)^k) + z^k \Psi(\{0^k (k-h)^k\}).
\label{eq:A7}
\end{equation}
Substituting for $\Psi(\{0^j (k-h)^k 0^{k-j}\})$ in Eq.~(\ref{eq:A5}) 
from Eq.~(\ref{eq:A6}), we obtain
\begin{equation}
\Psi(\{h^j 0^k h^{k-j}\} =  \frac{z}{(1 - z^{2k})} \Psi(\{(h-1)^j 0^k 
(h-1)^{k-j}\},
\label{eq:A8}
\end{equation}
which is immediately solved to give
\begin{equation}
\Psi(\{h^j 0^k h^{k-j}\}) = \frac{ z^h}{(1 - z^{2k})^h} \Psi(\{0^{2k}\}),
~1 \leq h,j  \leq k-1.
\label{eq:A9}
\end{equation}

Substituting for  $\Psi(\{(k-1)^j 0^k (k-1)^{k-j}\})$ in Eq.~(\ref{eq:A4}) from Eq.~(\ref{eq:A9}) and simplifying, we obtain
\bea
&&\left[ 1 - z^2 - z^{2k} -   (k-1) z^{2k} ( 1 - z^{2k})^{-k +1} \right] \Psi(\{0^{2k}\}) \nonumber \\
&&= 1 + 2 z^{k+1} \Psi(\{0^k (k-1)^k\}).
\label{eq:A10}
\eea
To close the equations, we have to determine $\Psi(\{0^k h^k\}) $ in 
terms of $\Psi(\{0^{2k}\})$. The values of $\Psi(\{0^k h^k\})$ for one 
value of $h$ are related by Eq.~(\ref{eq:A7}) to arguments $(h-1)$ and to 
$(k-h)$. This seems complicated, but it is easily checked that the 
ansatz 
\begin{equation}
\Psi(\{0^k h^k\}) = C \alpha^h +  D \alpha^{-h}, {\rm ~~for}~~ 0 \leq h 
\leq k-1.
\label{eq:A11}
\end{equation}
satisfies Eq.~(\ref{eq:A7}), so long as 
\begin{eqnarray}
(1 - \frac{z}{\alpha} ) C &=&  z^k \alpha^{-k} D, \nonumber\\
(1 - z \alpha) D &=&  z^k \alpha^{k} C.
\label{eq:A12}
\end{eqnarray}

Eliminating $C/D$ from Eq~(\ref{eq:A12}), we obtain
\begin{equation}
( 1 - \alpha z) ( 1 - \frac{z}{\alpha} ) = z^{2k}.
\label{eq:A13}
\end{equation}
This is a quadratic equation in  $\alpha$, and determines $\alpha$ for 
any given value of $z$.   Explicitly, we obtain
\begin{equation}
\alpha^{\pm 1} \!=\!\frac{1 +z^2 - z^{2k} \pm \sqrt{ \left[ (1 - z)^2 
-z^{2k} \right] \left[(1 +z)^2 - 
z^{2k}\right]}}{2z}.
\label{eq:A14}
\end{equation}

Using Eq.~(\ref{eq:A12}), we can express $C$ and $D$ in terms of a single 
variable $\kappa$ :
\begin{eqnarray}
C = \kappa \alpha^{-k/2} \sqrt{1 - z \alpha}, \label{eq:a15}\\
D = \kappa \alpha^{k/2}  \sqrt{1 - z /\alpha}. \label{eq:a16}
\end{eqnarray}
The actual values of $C$
and $D$ can be determined from the boundary condition at $h=0$:
\begin{equation}
C + D = \Psi(\{0^{2k}\}).
\label{eq:A17}
\end{equation}
We obtain
\begin{equation}
\kappa= \frac{\Psi(\{0^{2k}\})}{\alpha^{-k/2} (1 - z \alpha)^{1/2} + \alpha^{k/2} 
( 1 - z/ \alpha)^{1/2} }.
\label{eq:A18}
\end{equation}
Substituting for $\kappa$ in 
Eqs.~(\ref{eq:a15}) and (\ref{eq:a16}), we obtain $C$ and $D$ in terms of $z$. 
\begin{eqnarray}
C = \frac{ 1 - z \alpha }{ 1 - z \alpha + z^k \alpha^k} 
\Psi(\{0^{2k}\}),\\
D = \frac{ \alpha^k z^k}{1 - z \alpha + z^k \alpha^k} \Psi(\{0^{2k}\}).
\end{eqnarray}

Finally, substituting the value of $C$ and $D$ in Eq.~(\ref{eq:A11}), we 
obtain
\begin{equation}
\psi(\{0^k (k-1)^k\}) = \frac{ ( 1- z \alpha) \alpha^{k-1} + z^k 
\alpha }{1 - z \alpha + z^k \alpha^k} 
\Psi(\{0^{2k}\}).
\label{eq:A19}
\end{equation}
Equation~(\ref{eq:A19}) may be simplified by substituting for $z^k$ from Eq.~(\ref{eq:A13}):
\bea
&&\Psi(\{0^k(k-1)^k\}) = \label{eq:A22}\\
&& \Psi(\{0^{2k}\}) \left[\frac{ \alpha^{k/2-1} \sqrt{1 - 
z 
\alpha}+ \alpha^{-k/2 +1} \sqrt{ 1 - z/ \alpha}}{  \alpha^{-k/2} \sqrt{1 - z 
\alpha}+ \alpha^{k/2} 
\sqrt{1 - z/ \alpha}}\right]. \nonumber
\eea
Note the explicit symmetry of the expression under the exchange of $\alpha \leftrightarrow  1/\alpha$.

Substituting the expressions for $\alpha$, $\Psi(\{0^k(k-1)^k\})$ in Eq.~(\ref{eq:A10}), we obtain  an 
explicit expression for $\Psi(\{0^{2k}\})$ of the form  
\be
\Psi(\{0^{2k}\})= \frac{1}{E(z)},
\ee
where the denominator $E(z)$ equals
\bea
&&E(z) =1 - z^2 - z^{2k} -  \frac{ (k-1) z^{2k}}{  (1 - z^{2k})^{k- 1}} \nonumber\\
&&- 2 z^{k+1} \left[\frac{ \alpha^{k/2-1} \sqrt{1 - 
z 
\alpha}+ \alpha^{-k/2 +1} \sqrt{ 1 - z/ \alpha}}{  \alpha^{-k/2} \sqrt{1 - z 
\alpha}+ \alpha^{k/2} 
\sqrt{1 - z/ \alpha}}\right]
\eea

The entropy $S_{2k \times \infty}$ is given by $-k^{-1} \ln z_{2k}^*$, where $z_{2k}^*$ is the singularity of $\Psi(\{0^{2k}\})$ that is closest to the origin. We will show below that asymptotic behavior of entropy for strips of width $2k$  is the same as that of strips of width $k$. The explicit values of the entropies for strips $k \times \infty$ and $2 k \times \infty$ for $k$ upto $2^{31}$ are given in Table~\ref{table:table}, and compared with the asymptotic result  $k^{-2} \ln k$ in Eq.~(\ref{eq:sd}).
\begin{table}
\caption{\label{table:table}Entropy $S$ for full packing of rods of length $k$ on strips $k \times \infty$ and $2 k \times \infty$, compared with the leading asymptotic result $k^{-2} \ln k$ in Eq.~(\ref{eq:sd}).}
\begin{ruledtabular}
\begin{tabular}{cccc}
$k$ & $S_{k \times \infty}$ & $S_{2 k \times \infty}$ &$k^{-2} \ln k$\\
\hline
$2^1 $&$2.406059 \times 10^{-1} $&$2.609982 \times 10^{-1} $ & $1.732868 \times 10^{-1}$\\
$2^{3} $&$2.608540 \times 10^{-2} $&$2.929916\times 10^{-2} $ & $3.249127 \times 10^{-2}$ \\
$2^{5} $&$2.503880\times 10^{-3} $&$2.797511\times 10^{-3} $ & $3.384508 \times 10^{-3}$ \\
$2^{7} $&$2.190142 \times 10^{-4} $&$ 2.429123 \times 10^{-4} $ &$2.961444 \times 10^{-4}$\\
$2^{9} $&$1.791444 \times 10^{-5} $&$1.975853 \times 10^{-5} $ & $2.379732 \times10^{-05}$\\
$2^{11} $&$1.396705\times 10^{-6} $&$1.532938\times 10^{-6} $ & $1.817851 \times10^{-06}$\\
$2^{13} $&$1.051617\times 10^{-7} $&$1.148760\times 10^{-7} $ & $1.342731 \times10^{-07}$ \\
$2^{15} $&$7.714252\times 10^{-9} $&$8.388809\times 10^{-9} $ & $9.683154 \times10^{-09}$\\
$2^{17} $&$5.546713\times 10^{-10} $&$6.006134\times 10^{-10} $ & $6.858901 \times10^{-10}$\\
$2^{19} $&$3.925774\times 10^{-11} $&$4.234237\times 10^{-11} $ & $4.791144 \times10^{-11}$\\
$2^{21} $&$2.743428\times 10^{-12} $&$2.948320\times 10^{-12} $ & $3.309672 \times10^{-12}$\\
$2^{23} $&$1.897264\times 10^{-13} $&$2.032237\times 10^{-13} $ & $2.265549 \times10^{-13}$\\
$2^{25} $&$1.300702\times 10^{-14} $&$1.389037\times 10^{-14} $ &$1.539096 \times10^{-14}$\\
$2^{27} $&$8.851686\times 10^{-16} $&$9.426769\times 10^{-16} $ &$1.038890 \times10^{-15}$\\
$2^{29} $&$5.985929\times 10^{-17} $&$6.358717\times 10^{-17} $ &$6.974028 \times10^{-17}$\\
$2^{31} $&$4.025907\times 10^{-18} $&$4.266698\times 10^{-18} $ & $4.659372 \times10^{-18}$
\end{tabular}
\end{ruledtabular}
\end{table}

We now determine the leading singularity $z_{2k}^*$ of $\Psi(\{0^{2k}\})$ in the limit $k \gg 1$. To do so, consider the denominator $E(z)$. It has a square root singularity at $z_c$ when the discriminant in Eq.~(\ref{eq:A14}) equals zero.  By factorising the discriminant and writing in terms of the modulus of $z_c$, we obtain that $z_c$ satisfies the equation
\be
1-z_c-z_c^k = 0,
\label{eq:A25}
\ee
identical to that satisfied by $z^*$ for the strip $k \times \infty$ [see Eq.~(\ref{eq2}) with $\lambda=1/z$]. For large $k$, it has the solution
\be
\ln z_c = -\frac{\ln k}{k} \left(1- \frac{\ln \ln k}{\ln k}+ \ldots \right), ~~k \gg 1.
\label{eq:eq36}
\ee

We now show that $E(z)$ has a zero at $z_{2k}^* \lesssim z_c$. Following a bit of algebra, it can be shown that
\be
E(z_c) = \frac{-(k-1) z_c (1-z_c)}{(2-z_c)^{k-1} } \approx \frac{-\ln k}{k},~~ k \gg 1.
\ee
Clearly, $E(z_c) <0$, as $z_c<1$. At the same time, it is clear that $E(0)=1$ since $\Psi(\{0^{2k}\})|_{z=0}=1$. Therefore, there must be a zero $z_{2k}^*$ in $(0, z_c)$, leading to a  higher entropy for strips $2 k \times \infty$, as  evident in Table~\ref{table:table}. Also as $E(z_c) \to 0$ for large $k$, we expect that $z_{2k}^* \to z_c$ for large $k$.

We now show that the leading behavior of $z_{2k}^*$ and $z_c$ are identical for large $k$. We look for solutions 
\be
z_{2k}^* = e^{-\eta/k},
\label{eq:eq38}
\ee
where  $\eta/k^\theta \to 0$ for any $\theta>0$, and $\eta \to \infty$ for $k \gg 1$. In this limit, $\alpha \approx 1+ \sqrt{4 \eta/k}$, and after some algebra, $E(z^*)$ may be simplified to give
\be
E(z_{2k}^*) \approx \frac{2 \eta}{k} - k e^{-2 \eta} - 2 e^{-\eta}.
\ee
Equating $E(z_{2k}^*)$ to zero, we obtain
\be
\frac{2 \eta}{k} = k e^{-2 \eta} +2 e^{-\eta}.
\label{eq:A30}
\ee
From direct substitution,  it is straightforward to check that Eq.~(\ref{eq:A30}) is satisfied to leading order by $\eta=\ln k [1-\ln(\ln k)/(2\ln k]$. Equation~(\ref{eq:eq38}) then gives
\be
\ln z_{2k}^* = -\frac{\ln k}{k} \left[ 1 - \frac{\ln \ln k}{2\ln k} +\mathrm{lower~ order ~terms}\right],
\ee
Since the entropy is given by $-\ln z_{2k}^*/k$, we obtain
\be
S_{2k\times \infty} = \frac{\ln k}{k^2} \left(1-\frac{\ln\ln k}{2\ln k} + \ldots\right).
\label{eq:43}
\ee
The leading behaviour of $S_{2k\times \infty}$ coincides with that of $S_{k\times \infty}$ [see Eq.~(\ref{eq:10})]. The subleading term is different.
We thus obtain the same lower bound as given in Eq.~(\ref{eq4}).

\subsection{\label{subsection:Lxinfinity} Entropy of strips $L \times \infty$}

For general $L \times \infty$, though one can write down the recursion relations obeyed by different generating functions, it is not possible to find a closed form solution for them. Here, we provide an alternate analysis by determining a lower bound for the asymptotic behaviour of entropy of $L \times \infty$ strips, which will happen to coincide with the exact results for the strips $k \times \infty$ and $2 k \times \infty$

We define the generating function 
\begin{equation}
\Omega_L(z) = \sum_{M=0}^{\infty} N(L,M) z^{LM/k},
\end{equation}
with $N(L,0)=1$, by convention. Then, by direct enumeration,
\begin{eqnarray}
\Omega_1(z)&=& 1 + z + z^2 + z^3 + \ldots,\\
\Omega_2(z) &=&  1 +z^2 + z^4 + z^6 + \ldots,  ~k>2.
\end{eqnarray}
$\Omega_L(z)$ is sum of weights of all configurations of rods on a semi-infinite strip of width $L$, with the weight of a configuration of $n$ rods being $z^n$.  We have the further constraint that all rods must lie fully in a rectangular region, with both bottom and top edge horizontal, and no uncovered regions within the rectangle.
As a slightly more complicated example,  it is easily seen that $\Omega_k(z)$ is the generating function for $k \times \infty$ strips, and hence,
\begin{equation}
\Omega_k(z) = \frac{1}{1 - z -z^k}.
\label{eq:k-strip}
\end{equation}

\begin{figure}
\begin{center} 
\includegraphics[width = \columnwidth]{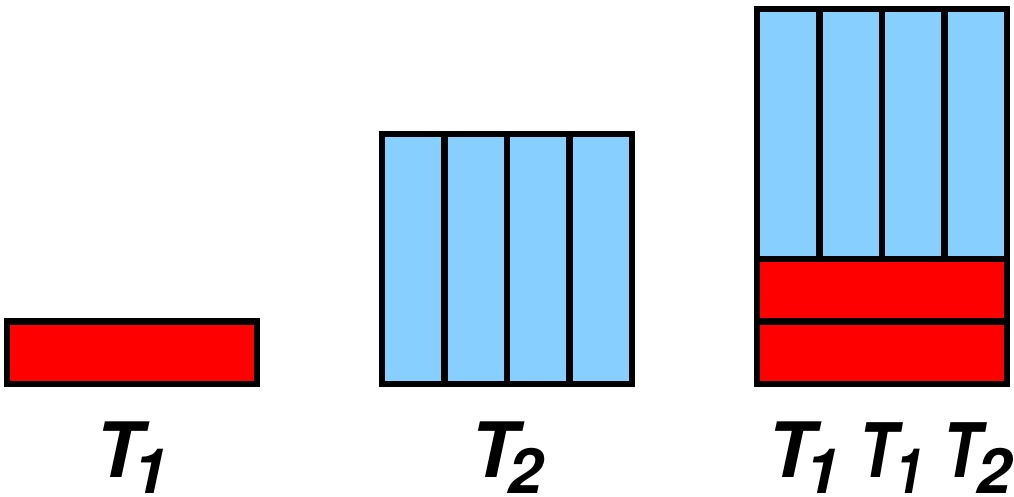}
\caption{\label{fig:concatenation}  A schematic diagram illustrating the procedure of vertical concatenation. Here $T_1$ and $T_2$ are two admissible tilings, and the  concatenated tiling $T_1 T_1 T_2$ is obtained by putting  two $T_1$s and one $T_2$  from bottom to top in the specified order.
}
\end{center} 
\end{figure}

In determining admissible tilings, a useful concept is that of {\it concatenation}. Given two tilings of  rectangles of sizes $L \times M_1$ and $L \times M_2$, we  define the vertical concatenation of these tilings as the tiling of size $L\times (M_1 + M_2)$, obtained by  just putting the  rectangles on top of each other in the order of concatenation.  An illustration of vertical concatenation of three tilings is shown in Fig.~\ref{fig:concatenation}. A horizontal concatenation is defined similarly.

A tiling is said to be vertically indecomposable, if it cannot be expressed as a vertical concatenation of two admissible tilings. For a vertically decomposable tiling, there is a horizontal line that divides the rectangle into two smaller tilings, such that no rod crosses the horizontal line.  

We now define $R_L(z)$ as the sum of weights of all vertically indecomposable tilings of rectangles of width $L$. Clearly, $R_L(z)$ is a series in powers of $z$, with all coefficients as non-negative integers.  Then, we have 
\begin{equation}
\Omega_L(z) = \frac{1}{1 - R_L(z)}.
\end{equation} 
Let  the radius of convergence of the power series of $\Omega_L(z)$ be $z_L^*$.  Then,
\begin{equation}
R_L(z_L^*)=1.
\end{equation}
Since $R_L(z)$ is a series of positive coefficients, we may truncate the series at any order, and obtain an upper bound estimate of $z_L^*$, and hence of $z^*$,  the limit of $z_L^*$, for large $L$. This, in turn, will provide a lower bound for the entropy.

We will take $L$ to be a multiple of $k$, as these give the best bounds.  The simplest case is $L=k$. In this case, $R_k(z)$ is a finite polynomial, and we have 
\begin{equation}
R_k(z) = z + z^k.
\end{equation}
We see that this is consistent with Eq.~(\ref{eq:k-strip}).

Now, let us consider the more complicated case $L = 2 k$. In this case, $R_{2k}(z)$ is not a finite polynomial.  But it has an interesting structure: the lowest order term is $z^2$, corresponding to a configuration of two horizontal
k-mers side by side. But, then terms of order $z^4, z^6, \ldots$  are all zero, as the corresponding tilings are decomposable.  The first non-zero term is of order $z^{2k}$, which corresponds to configurations consisting of a plaquette of $k$ aligned horizontal  rods and $k$ vertical rods tiling the remaining area, and another  of $2k$ vertical $k$-mers. With a small amount of brute force enumeration, it is easily seen that the plaquette can be placed in $(k+1)$ ways to give
\begin{equation}
R_{2k}(z) = z^2 + ( k + 2) z^{2k} + {\cal O}(z^{2k+2}).
\end{equation}

If we truncate the equation at order $z^{2k}$, we obtain an upper bound estimate for  $z_{2k}^*$. It turns out that for large $k$, the  terms that have been dropped make only a negligible  contribution to $R_{2k}(z)$ at $z=z_{2k}^*$. We will verify this claim  later. First, we solve the truncated equation for $z_{2k}^*$:
\begin{equation}
z_{2k}^{*2} + ( k+2) z_{2k}^{*2k} =1.
\end{equation}
Writing $z_{2k}^* = \exp(-B/k)$, we see that $(1 -z_{2k}^{*2}) \approx 2B/k$, if $k$ is large, to leading order in $k$,  $B$ satisfies the equation
\begin{equation}
2B \exp(2B)  \approx k ( k+2), 
\end{equation}
which has the solution $B \approx (1/2)W(k(k+2))$ [$W(z)$ being the Lambert function], which for large $k$ has the leading behavior 
\begin{equation}
B \approx \ln \left[ \sqrt  \frac {k (k +2)}{ 2 \ln k}\right]  \approx \ln \left(\frac{ k}{\sqrt{ 2 \ln k}} \right).
\end{equation}
Since $S_{2k \times \infty}\geq -\ln(z_{2lk}^*)/k$, we obtain
\be
S_{2k \times \infty} \geq  \frac{\ln k}{k^2} \left(1-\frac{\ln \ln k}{2\ln k} + \ldots \right).
\ee
This is a bit larger than the estimate using strips of width $k$ [see Eq.~(\ref{eq:10})], but for large $k$, the leading behavior remains the same with the difference showing only in the sub-leading correction of order $(\ln \ln k)/k$. We also note that the bound for $S_{2k \times \infty}$ obtained by truncation coincides with the exact analysis [see Eq.~(\ref{eq:43})].

Now, the term of order $z^{2k+2}$ in $R_{2k}$ is only $2 z^{2k+2}$, and its contribution to sum is smaller than that of the term of order $z^{2k}$ by a factor $(1/k)$.  At higher orders, the term of order $z^{6k}$ has a coefficient of order $k^3$. Using the fact that $z^k$ is of order $ 1/k$, the net contribution of this term decreases as  $1/k^3$. This also does not change the leading order  $k$-dependence of $z_{2k}^*$.

\begin{figure}
\begin{center} 
\includegraphics[width = \columnwidth]{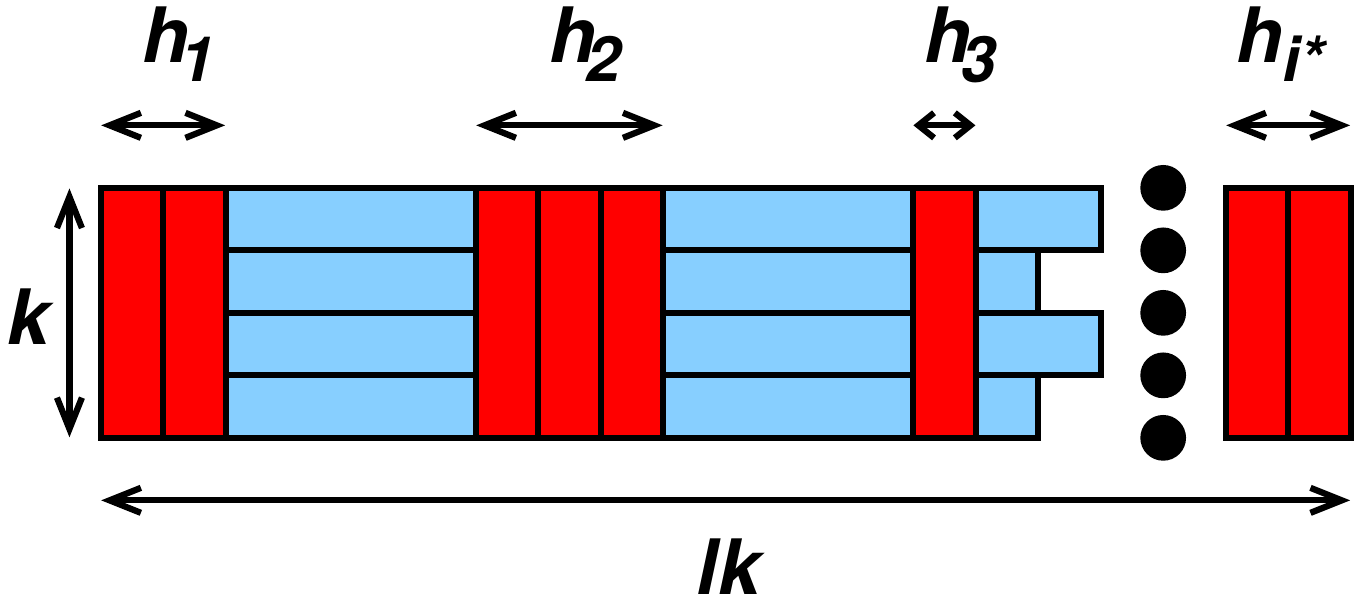}
\caption{\label{fig:fig9} A schematic representation of configurations that contribute to $C_l$ to leading order in $k$. Here, there must be exactly $k$ vertical $k$-mers so that $h_1 + h_2 +\ldots+h_{i^*}= k$. }
\end{center} 
\end{figure}

A similar argument works for other values of $R_{lk}(z)$.  We will only sketch the arguments here. The series expansion  for $R_{lk}(z)$ in powers of $z$ is of the form
\begin{equation}
R_{lk}(z) = z^l + C_l z^{lk} + {\rm ~higher~powers~of~} z^l.
\end{equation}
Here $C_l$ is an $l$-dependent coefficient.  The leading contribution to this term comes from  configurations  depicted in Fig.~\ref{fig:fig9}, consisting of  of $(l-1)$  plaquettes of $k$ aligned horizontal $k$-mers, interspersed with $k$ vertical rods.  The number of such configurations is $\dbinom{k+l-1}{l-1}$.  Thus, keeping only the first two non-trivial terms in the expansion for $R_l(z)$, we write 
 \begin{equation}
z_{lk}^{*l}  + \dbinom{k+l-1}{l-1} z_{lk}^{*kl} =1.
\end{equation}
Solving this equation, we see that its smallest positive  root has the leading $k$-dependence given by
\begin{equation}
z^*_{lk}  \approx \exp\left[ - \frac{1}{k} \ln  \frac {k}{ {(l! \ln k)^{1/l}}}\right],~~k \gg m.
\end{equation}
Since the entropy $S_{lk \times \infty} =  -\ln(z_{lk}^{*})/k$,  we obtain that
\be
S_2(k) \geq S_{lk \times \infty} \geq \frac{\ln k}{k^2} \left(1-\frac{\ln \ln k}{l \ln k} + \ldots \right), ~~k \gg 1.
\label{eq:59}
\ee
We conclude that
\be
\lim_{k\to \infty} \frac{k^2 S_2(k)}{\ln k} \geq 1. \label{eq:upperbound}
\ee

\section{\label{sec:upperbound} Upper bound for entropy for large $k$ and the main result}

Gagunashvili and Priezzhev obtained an upper bound for $N(L,M)$~\cite{gagunashvili1979close}.  They considered a  subset of sites  of the square lattice whose coordinates are multiple of $k$, and assumed that we are given the configuration of $k$-mers that cover these sites. Then, they proved that there is at most one way to cover the remaining sites with $k$-mers.  Then, the number of coverings allowed is bounded from above by the number of ways the subset of sites can be covered by $k$-mers. But each of these can be covered in at most $2 k$ ways.  Since there are at most $N/k^2$ such sites, they obtain
\begin{equation}
N(L,M) \leq  (2k)^{LM/k^2}.
\label{eq:priezzhev}
\end{equation}
This implies that $S_2(k) \leq  \ln(2k)/ k^2$, or equivalently
\be
\lim_{k\to \infty} \frac{k^2 S_2(k)}{\ln k} \leq 1.
\label{eq:priezzhev10}
\ee

In fact, Gagunashvili and Priezzhev proved a stronger upper bound  which for large $k$ is  $ S_2(k) \leq k^{-2} \ln( \gamma k)$, where $\gamma= \exp(4 G/\pi)/2$, with $G$ being the Catalan's constant.  Numerically, $\gamma \approx 1.605$. However,  the weaker bound is adequate for our purpose here

We now combine the lower bound obtained for entropy in 
Eqs.~(\ref{eq4}) and (\ref{eq:upperbound}), and the upper bound obtained for entropy in Eq.~(\ref{eq:priezzhev10}). Since, these two bounds are the same, we conclude that the entropy for fully packed rods on a square lattice has the asymptotic behavior 
\begin{equation}
\lim_{ k \rightarrow \infty}  \frac{ k^2 S_2(k)}{\ln k}  = 1,
\end{equation}
as given in Eq.~(\ref{eq:sd}), thus proving our main result.

\begin{figure} 
\begin{center}
\includegraphics[width=\columnwidth]{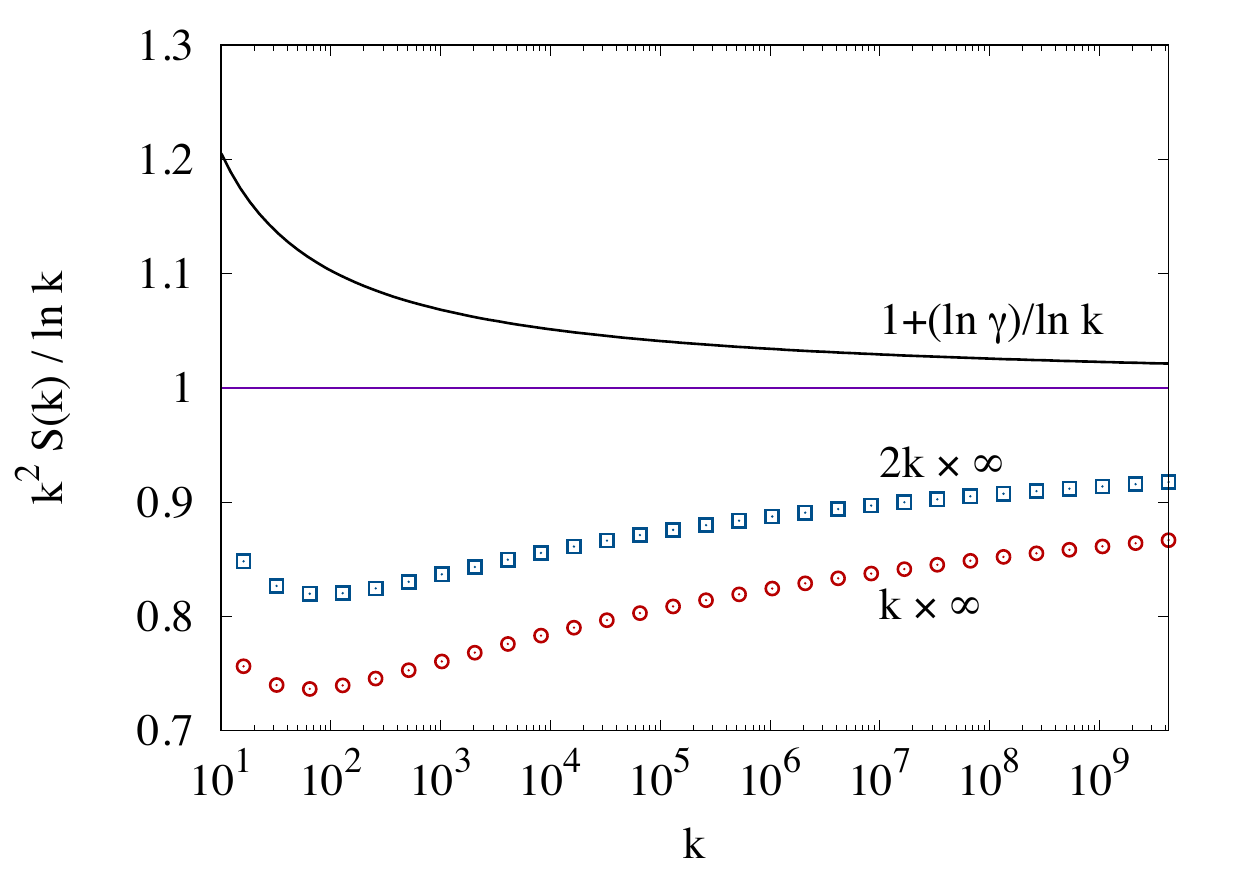}
\end{center}
\caption{\label{fig:fig12}  Bounds for the quantity $k^2 S_2(k)/\ln k$, whose asymptotic answer is one [see Eq.~(\ref{eq:sd})]. Lower bounds are provided by the entropies on strips $k \times \infty$ and $2 k \times \infty$. An upper bound~\cite{gagunashvili1979close} is provided by $1+\ln \gamma/\ln k$, where $\gamma\approx 1.605$. }
\end{figure}

We now look at how the bounds converge to the asymptotic result. The entropies on the strips $k \times \infty$ and $2 k \times \infty$ provide lower bounds for the entropy on infinite lattices. Ref.~\cite{gagunashvili1979close} gives an upper bound for large $k$ as $S_2(k) \leq k^{-2} \ln( \gamma k)$, where $\gamma \approx 1.605$. Since the leading form is the same for both the upper bounds as well as lower bounds, we divide it out by considering $k^2 S(k)/\ln k$, which converges to $1$ for large $k$. The strips provide lower bounds while $\ln \gamma $ provides an upper bound for this quantity. These bounds are shown in Fig.~\ref{fig:fig12}.

In addition, it is also possible to put a bound on the subleading corrections to the entropy. Let  $\beta$   be  the coefficient of the subleading term  $\ln(\ln k)/k^2$ in the asymptotic expansion of the entropy such that
\be
S_2(k) = \frac{\ln k}{k^2} + \beta \frac{\ln \ln k}{k^2} + \ldots.
\ee
From Eq.~(\ref{eq:59}), by taking the limit $l \to \infty$, it follows that
\be
\beta \geq 0.
\ee

\section{\label{sec:higherdim} Extension to higher dimensions}

We now present a heuristic argument that extends the above result  to   higher dimensions $ d > 2$. For a system of $k$-mers on a $d$-dimensional hypercubical lattice, with $ d>2$, we  argue below that configurations of rods that are fully layered  provide a  good starting point to calculate  the entropy of the system. In fact, this approach becomes better for larger $k$ and one can  develop a series expansion in the  number of rods that are between the layered planes.  Then, in a typical state, there are only a few such rods, and the full state will show spontaneous symmetry breaking, with most of the rods in the configuration being  one of the  $d \choose 2$ orientations.  In this way, one obtains the full packing constraint satisfied within a 2-dimensional layer, and different layers can be occupied independently, leading to a large entropy.  We note that the existence of layering in the high-density phase has been seen  in simulations at densities close to full packing for rods of length larger than or equal to five in $d=3$~\cite{2017-vdr-jsm-different}.

We first consider the case  $d=3$.  The argument is easily extended to higher $d$. 

We will use an adaption of the series expansion technique developed in the context of hard square lattice gases~\cite{1967-bn-jcp-phase,1966-bn-prl-phase,2012-rd-pre-high} and hard rectangle lattice gases~\cite{2015-nkr-jsp-high}.
For the simple cubic lattice, we consider  different activities $w_1, w_2, w_3$ for rods oriented along  the $x$-, $y$-, and $z$- directions. Let the corresponding partition function for an $L \times L \times L$ lattice be denoted by $\Omega_L( w_1,w_2,w_3)$. 

We will consider a perturbation expansion of this in powers of $w_3$.  We start with the case $w_1 = w_2 =w$, and $w_3 =0$.   The the grand  partition function  of a $L \times L \times L$ cuboid  $\Omega_L(w,w,0) $ can be written as  product of $2$-dimensional partition functions
\begin{equation}
\Omega_{L}(w,w,0)  = \left[\Omega_{2d,L}(w) \right]^{L},
\end{equation}
where $\Omega_{2d,L}(w)$ is the partition function of a full packing of $L \times L \times 1$ layer  by $k$-mers. 
We write $\Omega_L( w_1,w_2,w_3)$ as a perturbation expansion in $w_3$ about $w_3=0$. If this series expansion is well behaved, this implies  that for large enough $k$, the full packed configuration will show spontaneous symmetry breaking, and for large $k$, the fraction of non-planar rods in a random full planar configuration will tend to zero.  The first term in $w_3$ must be proportional to $L^d$ so that the entropy is extensive. We, thus, write the expansion as
\be
\Omega_L( w,w,w_3) = \Omega_L( w,w,0) \left[ 1+ L^d A_1 ~w_3^k + {\mathcal O}(w_3^{2k}) \right],
\label{eq:expansion1}
\ee
Taking logarithm and dividing by $L^d$, we obtain the expansion for the entropy:
\be
S_3(w,w, w_3) = S_3(w,w,0) + A_1  w_3^k + \ldots,
\ee
and we assume that this systematic expansion is well behaved, and  converges for small $w_3$.

The first nontrivial  term in Eq.~(\ref{eq:expansion1}) is proportional to $w_3^k$, and the coefficient  $A_1$ is determined in terms of  the number of configurations of  $k$ $z$-type rods (to be also called vertical rods), with the rest of the rods being of the $x$- and $y$- type. These vertical rods will have to be  in the same vertical slab of height $k$.  Let the $x$- and $y$-coordinates of the lowest point of these vertical rods $\{\alpha_i,\beta_i\}$, $i= 1$ to $k$. Let  the number of possible coverings   of rods in one plane, given unoccupied sites $\{x_i,y_i\}$, be $N(\{x_i,y_i\})$.  Then the number of coverings of the cuboid is proportional to $[N(\{x_i,y_i\})]^k$, and the relative weight of this  term will be $[N(\{\alpha_i,\beta_i\})/\Omega_{2d,L}]^k$. We note that  $[N(\{\alpha_i,\beta_i\})/\Omega_{2d,L}]^k$ may be considered as proportional to the probability distribution of  the bound state of $k$ holes. We then sum over different possible $\{\alpha_i,\beta_i\}$. We expect $N_{\alpha_i,\beta_i}/ \Omega_{2d,L}$ to be less than 1,  and to decrease as a power law of the distances between $\{\alpha_i , \beta_i\}$,  but with a power large enough so that 
\be
A_1=\sum^{\prime}_{\{\alpha_i,\beta_i\}}
\left[\frac{N(\{\alpha_i,\beta_i\})}{\Omega_{2d,L}}\right]^k < \infty,
\ee
where the primed sum is  over $i=2, 3, \ldots, k$, with  $\{\alpha_1, \beta_1\}$ fixed.  Then the sum over $\alpha_1$ and $\beta_1$ gives a factor proportional to $L^3$.

This is a complicated problem, for which we do not know the exact closed form  expression. However, we note that each term decreases exponentially with $k$ for large $k$ since we expect $N_{\alpha_i,\beta_i}/ \Omega_{2d,L}<1$. We note that $N(\{\alpha_i,\beta_i\})$ would be expected to be largest, when the holes are near each other.  In fact, the closest they can be be is in a continuous  single line of $k$ points, which may be created by removing  a single rod from the $2d$-covering. In this case, the contribution of the term is $[1/(2k)]^{k-1}$. 

If we sum to all orders, the dominant contribution will be expected to be of the same form.  We thus conclude that
for large $k$,  
\begin{equation}
S_3(k) = S_2(k) + {\rm ~terms~of~order~} k^{-k}.
\end{equation}
Then $\Omega_L(w_3)$,  as a function of $w_3$, and expand in powers of $w_3^k$, order by order, each term in the perturbation series will give an exponentially small  contribution in the large-$k$ limit.

Taking derivative   of $\ln \Omega_L(w,w,w_3)$  with respect to $\ln w_3$, we obtain
the fractional number of rods in the $z$-direction at full packing in this ensemble, and we see that this fraction tends to zero as $k$ increases, and the  departure
from perfect layering decreases for larger $k$.

The argument is immediately extended to higher $d$, and leads us to the conjecture in Eq.~(\ref{eq:sd}).

\section{\label{sec:conclusions} Concluding remarks}

In this paper, we studied the tiling of a  finite  rectangular part of the plane by rectangles of size $k \times 1$ and $1\times k$.  We showed that in order to get a full coverage, one of the sides of the rectangle to be covered should a multiple of $k$. We also studied the structure of the tilings of rectangles, and showed that all tilings can be obtained from each other by a sequence of basic flip move that exchanges a small $k \times k$ small square made of $k$ parallel vertical rectangles into horizontal ones, and vice versa.  We also showed provided non-rigorous perturbation theory based arguments for the conjecture that $S_d(k)$, the entropy per site  for $k$-mers on  a $d$-dimensional hypercubical lattice covered by straight rods of length $k$, for all $d \geq 2$  satisfies
Eq.~(\ref{eq:sd}). We emphasize that while the perturbation theory argument seems quite plausible, there is no proof that such a perturbation expansion is convergent.  If the series expansion does not converge,  or converges to a wrong value, the argument given here would break down.
 
The fact that this limit is independent of dimension deserves some comment. In general, we would expect the coefficients of logarithms encountered in the study of critical phenomena to be `universal', because by definition, they do not change  under a change of length scale in a renormalization transformation.  But, such coefficients are in general  not  dimension independent. In fact, here, $S_d$ has a multiplying factor $k^{-2}$, which indicates that the relevant quantity is the number of allowed configurations per unit square  of  length $k$ (which is the natural length scale in the problem). This number is proportional to $k$, and the entropy is proportional to $\ln k$. The fact that this is independent of dimension is only reflecting the fact that for large $k$, the problem essentially reduces a two-dimensional problem, because of spontaneous symmetry breaking, and most of the configurations at full packing are the ones where  the system breaks up into disjoint two-dimensional  layers. Consequently,  for large $k$, the leading behavior of entropy in higher dimensions is same as the two-dimensional case.

\section{Acknowledgments}
DD's  work was partially supported by the grant DST-SR-S2/JCB-24/2005 of the Government of India.

%\bibliography{ref}

\begin{thebibliography}{72}%
\makeatletter
\providecommand \@ifxundefined [1]{%
 \@ifx{#1\undefined}
}%
\providecommand \@ifnum [1]{%
 \ifnum #1\expandafter \@firstoftwo
 \else \expandafter \@secondoftwo
 \fi
}%
\providecommand \@ifx [1]{%
 \ifx #1\expandafter \@firstoftwo
 \else \expandafter \@secondoftwo
 \fi
}%
\providecommand \natexlab [1]{#1}%
\providecommand \enquote  [1]{``#1''}%
\providecommand \bibnamefont  [1]{#1}%
\providecommand \bibfnamefont [1]{#1}%
\providecommand \citenamefont [1]{#1}%
\providecommand \href@noop [0]{\@secondoftwo}%
\providecommand \href [0]{\begingroup \@sanitize@url \@href}%
\providecommand \@href[1]{\@@startlink{#1}\@@href}%
\providecommand \@@href[1]{\endgroup#1\@@endlink}%
\providecommand \@sanitize@url [0]{\catcode `\\12\catcode `\$12\catcode
  `\&12\catcode `\#12\catcode `\^12\catcode `\_12\catcode `\%12\relax}%
\providecommand \@@startlink[1]{}%
\providecommand \@@endlink[0]{}%
\providecommand \url  [0]{\begingroup\@sanitize@url \@url }%
\providecommand \@url [1]{\endgroup\@href {#1}{\urlprefix }}%
\providecommand \urlprefix  [0]{URL }%
\providecommand \Eprint [0]{\href }%
\providecommand \doibase [0]{https://doi.org/}%
\providecommand \selectlanguage [0]{\@gobble}%
\providecommand \bibinfo  [0]{\@secondoftwo}%
\providecommand \bibfield  [0]{\@secondoftwo}%
\providecommand \translation [1]{[#1]}%
\providecommand \BibitemOpen [0]{}%
\providecommand \bibitemStop [0]{}%
\providecommand \bibitemNoStop [0]{.\EOS\space}%
\providecommand \EOS [0]{\spacefactor3000\relax}%
\providecommand \BibitemShut  [1]{\csname bibitem#1\endcsname}%
\let\auto@bib@innerbib\@empty
%</preamble>
\bibitem [{\citenamefont {Alder}\ and\ \citenamefont
  {Wainwright}(1957)}]{1957-aw-jcp-phase}%
  \BibitemOpen
  \bibfield  {author} {\bibinfo {author} {\bibfnamefont {B.~J.}\ \bibnamefont
  {Alder}}\ and\ \bibinfo {author} {\bibfnamefont {T.~E.}\ \bibnamefont
  {Wainwright}},\ }\bibfield  {title} {\bibinfo {title} {Phase transition for a
  hard sphere system},\ }\href {https://doi.org/10.1063/1.1743957} {\bibfield
  {journal} {\bibinfo  {journal} {J. Chem. Phys.}\ }\textbf {\bibinfo {volume}
  {27}},\ \bibinfo {pages} {1208} (\bibinfo {year} {1957})}\BibitemShut
  {NoStop}%
\bibitem [{\citenamefont {Alder}\ and\ \citenamefont
  {Wainwright}(1962)}]{1962-aw-pr-phase}%
  \BibitemOpen
  \bibfield  {author} {\bibinfo {author} {\bibfnamefont {B.~J.}\ \bibnamefont
  {Alder}}\ and\ \bibinfo {author} {\bibfnamefont {T.~E.}\ \bibnamefont
  {Wainwright}},\ }\bibfield  {title} {\bibinfo {title} {Phase transition in
  elastic disks},\ }\href {https://doi.org/10.1103/PhysRev.127.359} {\bibfield
  {journal} {\bibinfo  {journal} {Phys. Rev.}\ }\textbf {\bibinfo {volume}
  {127}},\ \bibinfo {pages} {359} (\bibinfo {year} {1962})}\BibitemShut
  {NoStop}%
\bibitem [{\citenamefont {Noya}\ \emph {et~al.}(2008)\citenamefont {Noya},
  \citenamefont {Vega},\ and\ \citenamefont
  {de~Miguel}}]{noya2008determination}%
  \BibitemOpen
  \bibfield  {author} {\bibinfo {author} {\bibfnamefont {E.~G.}\ \bibnamefont
  {Noya}}, \bibinfo {author} {\bibfnamefont {C.}~\bibnamefont {Vega}},\ and\
  \bibinfo {author} {\bibfnamefont {E.}~\bibnamefont {de~Miguel}},\ }\bibfield
  {title} {\bibinfo {title} {Determination of the melting point of hard spheres
  from direct coexistence simulation methods},\ }\href@noop {} {\bibfield
  {journal} {\bibinfo  {journal} {J. Chem. Phys.}\ }\textbf {\bibinfo {volume}
  {128}},\ \bibinfo {pages} {154507} (\bibinfo {year} {2008})}\BibitemShut
  {NoStop}%
\bibitem [{\citenamefont {Pusey}\ and\ \citenamefont {van
  Megen}(1986)}]{1986-pm-nature-phase}%
  \BibitemOpen
  \bibfield  {author} {\bibinfo {author} {\bibfnamefont {P.~N.}\ \bibnamefont
  {Pusey}}\ and\ \bibinfo {author} {\bibfnamefont {W.}~\bibnamefont {van
  Megen}},\ }\bibfield  {title} {\bibinfo {title} {Phase behaviour of
  concentrated suspensions of nearly hard colloidal spheres},\ }\href
  {https://www.nature.com/articles/320340a0} {\bibfield  {journal} {\bibinfo
  {journal} {Nature}\ }\textbf {\bibinfo {volume} {320}},\ \bibinfo {pages}
  {340} (\bibinfo {year} {1986})}\BibitemShut {NoStop}%
\bibitem [{\citenamefont {Fisher}(1966)}]{fisher1966dimer}%
  \BibitemOpen
  \bibfield  {author} {\bibinfo {author} {\bibfnamefont {M.~E.}\ \bibnamefont
  {Fisher}},\ }\bibfield  {title} {\bibinfo {title} {On the dimer solution of
  planar ising models},\ }\href@noop {} {\bibfield  {journal} {\bibinfo
  {journal} {J. Math. Phys.}\ }\textbf {\bibinfo {volume} {7}},\ \bibinfo
  {pages} {1776} (\bibinfo {year} {1966})}\BibitemShut {NoStop}%
\bibitem [{\citenamefont {Huse}\ \emph {et~al.}(2003)\citenamefont {Huse},
  \citenamefont {Krauth}, \citenamefont {Moessner},\ and\ \citenamefont
  {Sondhi}}]{2003-hkms-prl-coulomb}%
  \BibitemOpen
  \bibfield  {author} {\bibinfo {author} {\bibfnamefont {D.~A.}\ \bibnamefont
  {Huse}}, \bibinfo {author} {\bibfnamefont {W.}~\bibnamefont {Krauth}},
  \bibinfo {author} {\bibfnamefont {R.}~\bibnamefont {Moessner}},\ and\
  \bibinfo {author} {\bibfnamefont {S.~L.}\ \bibnamefont {Sondhi}},\ }\bibfield
   {title} {\bibinfo {title} {Coulomb and liquid dimer models in three
  dimensions},\ }\href {https://doi.org/10.1103/PhysRevLett.91.167004}
  {\bibfield  {journal} {\bibinfo  {journal} {Phys. Rev. Lett.}\ }\textbf
  {\bibinfo {volume} {91}},\ \bibinfo {pages} {167004} (\bibinfo {year}
  {2003})}\BibitemShut {NoStop}%
\bibitem [{\citenamefont {M{\"o}essner}\ and\ \citenamefont
  {Sondhi}(2003)}]{mossner2003ising}%
  \BibitemOpen
  \bibfield  {author} {\bibinfo {author} {\bibfnamefont {R.}~\bibnamefont
  {M{\"o}essner}}\ and\ \bibinfo {author} {\bibfnamefont {S.~L.}\ \bibnamefont
  {Sondhi}},\ }\bibfield  {title} {\bibinfo {title} {Ising and dimer models in
  two and three dimensions},\ }\href@noop {} {\bibfield  {journal} {\bibinfo
  {journal} {Phys. Rev. B}\ }\textbf {\bibinfo {volume} {68}},\ \bibinfo
  {pages} {054405} (\bibinfo {year} {2003})}\BibitemShut {NoStop}%
\bibitem [{\citenamefont {de~Gennes}\ and\ \citenamefont
  {Prost}(1995)}]{1995-oup-gp-physics}%
  \BibitemOpen
  \bibfield  {author} {\bibinfo {author} {\bibfnamefont {P.~G.}\ \bibnamefont
  {de~Gennes}}\ and\ \bibinfo {author} {\bibfnamefont {J.}~\bibnamefont
  {Prost}},\ }\href@noop {} {\emph {\bibinfo {title} {The physics of liquid
  crystals}}},\ Vol.~\bibinfo {volume} {83}\ (\bibinfo  {publisher} {Oxford
  university press},\ \bibinfo {year} {1995})\BibitemShut {NoStop}%
\bibitem [{\citenamefont {Frenkel}(1988)}]{frenkel1988structure}%
  \BibitemOpen
  \bibfield  {author} {\bibinfo {author} {\bibfnamefont {D.}~\bibnamefont
  {Frenkel}},\ }\bibfield  {title} {\bibinfo {title} {Structure of hard-core
  models for liquid crystals},\ }\href@noop {} {\bibfield  {journal} {\bibinfo
  {journal} {J. Phys. Chem.}\ }\textbf {\bibinfo {volume} {92}},\ \bibinfo
  {pages} {3280} (\bibinfo {year} {1988})}\BibitemShut {NoStop}%
\bibitem [{\citenamefont {Care}\ and\ \citenamefont
  {Cleaver}(2005)}]{care2005computer}%
  \BibitemOpen
  \bibfield  {author} {\bibinfo {author} {\bibfnamefont {C.}~\bibnamefont
  {Care}}\ and\ \bibinfo {author} {\bibfnamefont {D.}~\bibnamefont {Cleaver}},\
  }\bibfield  {title} {\bibinfo {title} {Computer simulation of liquid
  crystals},\ }\href@noop {} {\bibfield  {journal} {\bibinfo  {journal} {Rep.
  Prog. Phys.}\ }\textbf {\bibinfo {volume} {68}},\ \bibinfo {pages} {2665}
  (\bibinfo {year} {2005})}\BibitemShut {NoStop}%
\bibitem [{\citenamefont {Donev}\ \emph {et~al.}(2004)\citenamefont {Donev},
  \citenamefont {Torquato}, \citenamefont {Stillinger},\ and\ \citenamefont
  {Connelly}}]{donev2004jamming}%
  \BibitemOpen
  \bibfield  {author} {\bibinfo {author} {\bibfnamefont {A.}~\bibnamefont
  {Donev}}, \bibinfo {author} {\bibfnamefont {S.}~\bibnamefont {Torquato}},
  \bibinfo {author} {\bibfnamefont {F.~H.}\ \bibnamefont {Stillinger}},\ and\
  \bibinfo {author} {\bibfnamefont {R.}~\bibnamefont {Connelly}},\ }\bibfield
  {title} {\bibinfo {title} {Jamming in hard sphere and disk packings},\
  }\href@noop {} {\bibfield  {journal} {\bibinfo  {journal} {J. Appl. Phys.}\
  }\textbf {\bibinfo {volume} {95}},\ \bibinfo {pages} {989} (\bibinfo {year}
  {2004})}\BibitemShut {NoStop}%
\bibitem [{\citenamefont {Liggett}(2012)}]{liggett2012interacting}%
  \BibitemOpen
  \bibfield  {author} {\bibinfo {author} {\bibfnamefont {T.~M.}\ \bibnamefont
  {Liggett}},\ }\href@noop {} {\emph {\bibinfo {title} {Interacting particle
  systems}}},\ Vol.\ \bibinfo {volume} {276}\ (\bibinfo  {publisher} {Springer
  Science \& Business Media},\ \bibinfo {year} {2012})\BibitemShut {NoStop}%
\bibitem [{\citenamefont {Mallick}(2015)}]{mallick2015exclusion}%
  \BibitemOpen
  \bibfield  {author} {\bibinfo {author} {\bibfnamefont {K.}~\bibnamefont
  {Mallick}},\ }\bibfield  {title} {\bibinfo {title} {The exclusion process: A
  paradigm for non-equilibrium behaviour},\ }\href@noop {} {\bibfield
  {journal} {\bibinfo  {journal} {Physica A}\ }\textbf {\bibinfo {volume}
  {418}},\ \bibinfo {pages} {17} (\bibinfo {year} {2015})}\BibitemShut
  {NoStop}%
\bibitem [{\citenamefont {Bellemans}\ and\ \citenamefont
  {Nigam}(1966)}]{1966-bn-prl-phase}%
  \BibitemOpen
  \bibfield  {author} {\bibinfo {author} {\bibfnamefont {A.}~\bibnamefont
  {Bellemans}}\ and\ \bibinfo {author} {\bibfnamefont {R.~K.}\ \bibnamefont
  {Nigam}},\ }\bibfield  {title} {\bibinfo {title} {Phase transitions in the
  hard-square lattice gas},\ }\href
  {https://doi.org/10.1103/PhysRevLett.16.1038} {\bibfield  {journal} {\bibinfo
   {journal} {Phys. Rev. Lett.}\ }\textbf {\bibinfo {volume} {16}},\ \bibinfo
  {pages} {1038} (\bibinfo {year} {1966})}\BibitemShut {NoStop}%
\bibitem [{\citenamefont {Bellemans}\ and\ \citenamefont
  {Nigam}(1967)}]{1967-bn-jcp-phase}%
  \BibitemOpen
  \bibfield  {author} {\bibinfo {author} {\bibfnamefont {A.}~\bibnamefont
  {Bellemans}}\ and\ \bibinfo {author} {\bibfnamefont {R.~K.}\ \bibnamefont
  {Nigam}},\ }\bibfield  {title} {\bibinfo {title} {Phase transitions in
  two‐dimensional lattice gases of hard‐square molecules},\ }\href
  {https://doi.org/10.1063/1.1841157} {\bibfield  {journal} {\bibinfo
  {journal} {J. Chem. Phys.}\ }\textbf {\bibinfo {volume} {46}},\ \bibinfo
  {pages} {2922} (\bibinfo {year} {1967})}\BibitemShut {NoStop}%
\bibitem [{\citenamefont {Ramola}\ and\ \citenamefont
  {Dhar}(2012)}]{2012-rd-pre-high}%
  \BibitemOpen
  \bibfield  {author} {\bibinfo {author} {\bibfnamefont {K.}~\bibnamefont
  {Ramola}}\ and\ \bibinfo {author} {\bibfnamefont {D.}~\bibnamefont {Dhar}},\
  }\bibfield  {title} {\bibinfo {title} {High-activity perturbation expansion
  for the hard square lattice gas},\ }\href
  {https://doi.org/10.1103/PhysRevE.86.031135} {\bibfield  {journal} {\bibinfo
  {journal} {Phys. Rev. E}\ }\textbf {\bibinfo {volume} {86}},\ \bibinfo
  {pages} {031135} (\bibinfo {year} {2012})}\BibitemShut {NoStop}%
\bibitem [{\citenamefont {Ramola}\ \emph {et~al.}(2015)\citenamefont {Ramola},
  \citenamefont {Damle},\ and\ \citenamefont {Dhar}}]{2015-rdd-prl-columnar}%
  \BibitemOpen
  \bibfield  {author} {\bibinfo {author} {\bibfnamefont {K.}~\bibnamefont
  {Ramola}}, \bibinfo {author} {\bibfnamefont {K.}~\bibnamefont {Damle}},\ and\
  \bibinfo {author} {\bibfnamefont {D.}~\bibnamefont {Dhar}},\ }\bibfield
  {title} {\bibinfo {title} {Columnar order and ashkin-teller criticality in
  mixtures of hard squares and dimers},\ }\href
  {https://doi.org/10.1103/PhysRevLett.114.190601} {\bibfield  {journal}
  {\bibinfo  {journal} {Phys. Rev. Lett.}\ }\textbf {\bibinfo {volume} {114}},\
  \bibinfo {pages} {190601} (\bibinfo {year} {2015})}\BibitemShut {NoStop}%
\bibitem [{\citenamefont {Nath}\ \emph {et~al.}(2016)\citenamefont {Nath},
  \citenamefont {Dhar},\ and\ \citenamefont {Rajesh}}]{2016-ndr-epl-stability}%
  \BibitemOpen
  \bibfield  {author} {\bibinfo {author} {\bibfnamefont {T.}~\bibnamefont
  {Nath}}, \bibinfo {author} {\bibfnamefont {D.}~\bibnamefont {Dhar}},\ and\
  \bibinfo {author} {\bibfnamefont {R.}~\bibnamefont {Rajesh}},\ }\bibfield
  {title} {\bibinfo {title} {Stability of columnar order in assemblies of hard
  rectangles or squares},\ }\href
  {http://stacks.iop.org/0295-5075/114/i=1/a=10003} {\bibfield  {journal}
  {\bibinfo  {journal} {Eur. Phys. Lett.}\ }\textbf {\bibinfo {volume} {114}},\
  \bibinfo {pages} {10003} (\bibinfo {year} {2016})}\BibitemShut {NoStop}%
\bibitem [{\citenamefont {Mandal}\ \emph {et~al.}(2017)\citenamefont {Mandal},
  \citenamefont {Nath},\ and\ \citenamefont {Rajesh}}]{mandal2017estimating}%
  \BibitemOpen
  \bibfield  {author} {\bibinfo {author} {\bibfnamefont {D.}~\bibnamefont
  {Mandal}}, \bibinfo {author} {\bibfnamefont {T.}~\bibnamefont {Nath}},\ and\
  \bibinfo {author} {\bibfnamefont {R.}~\bibnamefont {Rajesh}},\ }\bibfield
  {title} {\bibinfo {title} {Estimating the critical parameters of the hard
  square lattice gas model},\ }\href@noop {} {\bibfield  {journal} {\bibinfo
  {journal} {J. Stat. Mech.}\ }\textbf {\bibinfo {volume} {2017}},\ \bibinfo
  {pages} {043201} (\bibinfo {year} {2017})}\BibitemShut {NoStop}%
\bibitem [{\citenamefont {Verberkmoes}\ and\ \citenamefont
  {Nienhuis}(1999)}]{1999-vn-prl-triangular}%
  \BibitemOpen
  \bibfield  {author} {\bibinfo {author} {\bibfnamefont {A.}~\bibnamefont
  {Verberkmoes}}\ and\ \bibinfo {author} {\bibfnamefont {B.}~\bibnamefont
  {Nienhuis}},\ }\bibfield  {title} {\bibinfo {title} {Triangular trimers on
  the triangular lattice: An exact solution},\ }\href
  {https://doi.org/10.1103/PhysRevLett.83.3986} {\bibfield  {journal} {\bibinfo
   {journal} {Phys. Rev. Lett.}\ }\textbf {\bibinfo {volume} {83}},\ \bibinfo
  {pages} {3986} (\bibinfo {year} {1999})}\BibitemShut {NoStop}%
\bibitem [{\citenamefont {Baxter}(1980)}]{1980-b-jpa-exact}%
  \BibitemOpen
  \bibfield  {author} {\bibinfo {author} {\bibfnamefont {R.~J.}\ \bibnamefont
  {Baxter}},\ }\bibfield  {title} {\bibinfo {title} {Hard hexagons: exact
  solution},\ }\href {http://stacks.iop.org/0305-4470/13/i=3/a=007} {\bibfield
  {journal} {\bibinfo  {journal} {J. Phys. A}\ }\textbf {\bibinfo {volume}
  {13}},\ \bibinfo {pages} {L61} (\bibinfo {year} {1980})}\BibitemShut
  {NoStop}%
\bibitem [{\citenamefont {Flory}(1956)}]{1956-f-prs-phase}%
  \BibitemOpen
  \bibfield  {author} {\bibinfo {author} {\bibfnamefont {P.~J.}\ \bibnamefont
  {Flory}},\ }\bibfield  {title} {\bibinfo {title} {Phase equilibria in
  solutions of rod-like particles},\ }\href
  {https://doi.org/10.1098/rspa.1956.0016} {\bibfield  {journal} {\bibinfo
  {journal} {Proc. Roy. Soc. A}\ }\textbf {\bibinfo {volume} {234}},\ \bibinfo
  {pages} {73} (\bibinfo {year} {1956})}\BibitemShut {NoStop}%
\bibitem [{\citenamefont {Ghosh}\ and\ \citenamefont
  {Dhar}(2007)}]{2007-gd-epl-on}%
  \BibitemOpen
  \bibfield  {author} {\bibinfo {author} {\bibfnamefont {A.}~\bibnamefont
  {Ghosh}}\ and\ \bibinfo {author} {\bibfnamefont {D.}~\bibnamefont {Dhar}},\
  }\bibfield  {title} {\bibinfo {title} {On the orientational ordering of long
  rods on a lattice},\ }\href {http://stacks.iop.org/0295-5075/78/i=2/a=20003}
  {\bibfield  {journal} {\bibinfo  {journal} {Eur. Phys. Lett.}\ }\textbf
  {\bibinfo {volume} {78}},\ \bibinfo {pages} {20003} (\bibinfo {year}
  {2007})}\BibitemShut {NoStop}%
\bibitem [{\citenamefont {Dhar}\ \emph {et~al.}(2011)\citenamefont {Dhar},
  \citenamefont {Rajesh},\ and\ \citenamefont {Stilck}}]{2011-drs-pre-hard}%
  \BibitemOpen
  \bibfield  {author} {\bibinfo {author} {\bibfnamefont {D.}~\bibnamefont
  {Dhar}}, \bibinfo {author} {\bibfnamefont {R.}~\bibnamefont {Rajesh}},\ and\
  \bibinfo {author} {\bibfnamefont {J.~F.}\ \bibnamefont {Stilck}},\ }\bibfield
   {title} {\bibinfo {title} {Hard rigid rods on a bethe-like lattice},\ }\href
  {https://doi.org/10.1103/PhysRevE.84.011140} {\bibfield  {journal} {\bibinfo
  {journal} {Phys. Rev. E}\ }\textbf {\bibinfo {volume} {84}},\ \bibinfo
  {pages} {011140} (\bibinfo {year} {2011})}\BibitemShut {NoStop}%
\bibitem [{\citenamefont {Kundu}\ and\ \citenamefont
  {Rajesh}(2014)}]{2014-kr-pre-phase}%
  \BibitemOpen
  \bibfield  {author} {\bibinfo {author} {\bibfnamefont {J.}~\bibnamefont
  {Kundu}}\ and\ \bibinfo {author} {\bibfnamefont {R.}~\bibnamefont {Rajesh}},\
  }\bibfield  {title} {\bibinfo {title} {Phase transitions in a system of hard
  rectangles on the square lattice},\ }\href
  {https://doi.org/10.1103/PhysRevE.89.052124} {\bibfield  {journal} {\bibinfo
  {journal} {Phys. Rev. E}\ }\textbf {\bibinfo {volume} {89}},\ \bibinfo
  {pages} {052124} (\bibinfo {year} {2014})}\BibitemShut {NoStop}%
\bibitem [{\citenamefont {Kundu}\ and\ \citenamefont
  {Rajesh}(2015{\natexlab{a}})}]{2015-kr-epjb-phase}%
  \BibitemOpen
  \bibfield  {author} {\bibinfo {author} {\bibfnamefont {J.}~\bibnamefont
  {Kundu}}\ and\ \bibinfo {author} {\bibfnamefont {R.}~\bibnamefont {Rajesh}},\
  }\bibfield  {title} {\bibinfo {title} {Phase transitions in systems of hard
  rectangles with non-integer aspect ratio},\ }\href
  {https://doi.org/10.1140/epjb/e2015-60210-7} {\bibfield  {journal} {\bibinfo
  {journal} {Eur. Phys. J. B}\ }\textbf {\bibinfo {volume} {88}},\ \bibinfo
  {pages} {133} (\bibinfo {year} {2015}{\natexlab{a}})}\BibitemShut {NoStop}%
\bibitem [{\citenamefont {Kundu}\ and\ \citenamefont
  {Rajesh}(2015{\natexlab{b}})}]{2015-kr-pre-asymptotic}%
  \BibitemOpen
  \bibfield  {author} {\bibinfo {author} {\bibfnamefont {J.}~\bibnamefont
  {Kundu}}\ and\ \bibinfo {author} {\bibfnamefont {R.}~\bibnamefont {Rajesh}},\
  }\bibfield  {title} {\bibinfo {title} {Asymptotic behavior of the
  isotropic-nematic and nematic-columnar phase boundaries for the system of
  hard rectangles on a square lattice},\ }\href
  {https://doi.org/10.1103/PhysRevE.91.012105} {\bibfield  {journal} {\bibinfo
  {journal} {Phys. Rev. E}\ }\textbf {\bibinfo {volume} {91}},\ \bibinfo
  {pages} {012105} (\bibinfo {year} {2015}{\natexlab{b}})}\BibitemShut
  {NoStop}%
\bibitem [{\citenamefont {Szabelski}\ \emph {et~al.}(2013)\citenamefont
  {Szabelski}, \citenamefont {R{\.z}ysko}, \citenamefont {Pa{\'n}czyk},
  \citenamefont {Ghijsens}, \citenamefont {Tahara}, \citenamefont {Tobe},\ and\
  \citenamefont {De~Feyter}}]{szabelski2013self}%
  \BibitemOpen
  \bibfield  {author} {\bibinfo {author} {\bibfnamefont {P.}~\bibnamefont
  {Szabelski}}, \bibinfo {author} {\bibfnamefont {W.}~\bibnamefont
  {R{\.z}ysko}}, \bibinfo {author} {\bibfnamefont {T.}~\bibnamefont
  {Pa{\'n}czyk}}, \bibinfo {author} {\bibfnamefont {E.}~\bibnamefont
  {Ghijsens}}, \bibinfo {author} {\bibfnamefont {K.}~\bibnamefont {Tahara}},
  \bibinfo {author} {\bibfnamefont {Y.}~\bibnamefont {Tobe}},\ and\ \bibinfo
  {author} {\bibfnamefont {S.}~\bibnamefont {De~Feyter}},\ }\bibfield  {title}
  {\bibinfo {title} {Self-assembly of molecular tripods in two dimensions:
  structure and thermodynamics from computer simulations},\ }\href@noop {}
  {\bibfield  {journal} {\bibinfo  {journal} {RSC Adv.}\ }\textbf {\bibinfo
  {volume} {3}},\ \bibinfo {pages} {25159} (\bibinfo {year}
  {2013})}\BibitemShut {NoStop}%
\bibitem [{\citenamefont {Ruth}\ \emph {et~al.}(2015)\citenamefont {Ruth},
  \citenamefont {Toral}, \citenamefont {Holz}, \citenamefont {Rickman},\ and\
  \citenamefont {Gunton}}]{2015-rthrg-tsf-impact}%
  \BibitemOpen
  \bibfield  {author} {\bibinfo {author} {\bibfnamefont {D.}~\bibnamefont
  {Ruth}}, \bibinfo {author} {\bibfnamefont {R.}~\bibnamefont {Toral}},
  \bibinfo {author} {\bibfnamefont {D.}~\bibnamefont {Holz}}, \bibinfo {author}
  {\bibfnamefont {J.}~\bibnamefont {Rickman}},\ and\ \bibinfo {author}
  {\bibfnamefont {J.}~\bibnamefont {Gunton}},\ }\bibfield  {title} {\bibinfo
  {title} {Impact of surface interactions on the phase behavior of y-shaped
  molecules},\ }\href
  {https://doi.org/https://doi.org/10.1016/j.tsf.2015.11.046} {\bibfield
  {journal} {\bibinfo  {journal} {Thin Solid Films}\ }\textbf {\bibinfo
  {volume} {597}},\ \bibinfo {pages} {188 } (\bibinfo {year}
  {2015})}\BibitemShut {NoStop}%
\bibitem [{\citenamefont {Mandal}\ \emph {et~al.}(2018)\citenamefont {Mandal},
  \citenamefont {Nath},\ and\ \citenamefont {Rajesh}}]{2018-pre-mnr-phase}%
  \BibitemOpen
  \bibfield  {author} {\bibinfo {author} {\bibfnamefont {D.}~\bibnamefont
  {Mandal}}, \bibinfo {author} {\bibfnamefont {T.}~\bibnamefont {Nath}},\ and\
  \bibinfo {author} {\bibfnamefont {R.}~\bibnamefont {Rajesh}},\ }\bibfield
  {title} {\bibinfo {title} {Phase transitions in a system of hard y-shaped
  particles on the triangular lattice},\ }\href
  {https://doi.org/10.1103/PhysRevE.97.032131} {\bibfield  {journal} {\bibinfo
  {journal} {Phys. Rev. E}\ }\textbf {\bibinfo {volume} {97}},\ \bibinfo
  {pages} {032131} (\bibinfo {year} {2018})}\BibitemShut {NoStop}%
\bibitem [{\citenamefont {Barnes}\ \emph {et~al.}(2009)\citenamefont {Barnes},
  \citenamefont {Siderius},\ and\ \citenamefont
  {Gelb}}]{2009-bsg-langmuir-structure}%
  \BibitemOpen
  \bibfield  {author} {\bibinfo {author} {\bibfnamefont {B.~C.}\ \bibnamefont
  {Barnes}}, \bibinfo {author} {\bibfnamefont {D.~W.}\ \bibnamefont
  {Siderius}},\ and\ \bibinfo {author} {\bibfnamefont {L.~D.}\ \bibnamefont
  {Gelb}},\ }\bibfield  {title} {\bibinfo {title} {Structure, thermodynamics,
  and solubility in tetromino fluids},\ }\href
  {https://doi.org/10.1021/la900196b} {\bibfield  {journal} {\bibinfo
  {journal} {Langmuir}\ }\textbf {\bibinfo {volume} {25}},\ \bibinfo {pages}
  {6702} (\bibinfo {year} {2009})}\BibitemShut {NoStop}%
\bibitem [{\citenamefont {Panagiotopoulos}(2005)}]{2005-p-jcp-thermodynamic}%
  \BibitemOpen
  \bibfield  {author} {\bibinfo {author} {\bibfnamefont {A.~Z.}\ \bibnamefont
  {Panagiotopoulos}},\ }\bibfield  {title} {\bibinfo {title} {Thermodynamic
  properties of lattice hard-sphere models},\ }\href
  {https://doi.org/10.1063/1.2008253} {\bibfield  {journal} {\bibinfo
  {journal} {J. Chem. Phys.}\ }\textbf {\bibinfo {volume} {123}},\ \bibinfo
  {pages} {104504} (\bibinfo {year} {2005})}\BibitemShut {NoStop}%
\bibitem [{\citenamefont {Fernandes}\ \emph {et~al.}(2007)\citenamefont
  {Fernandes}, \citenamefont {Arenzon},\ and\ \citenamefont
  {Levin}}]{2007-fal-jcp-monte}%
  \BibitemOpen
  \bibfield  {author} {\bibinfo {author} {\bibfnamefont {H.~C.~M.}\
  \bibnamefont {Fernandes}}, \bibinfo {author} {\bibfnamefont {J.~J.}\
  \bibnamefont {Arenzon}},\ and\ \bibinfo {author} {\bibfnamefont
  {Y.}~\bibnamefont {Levin}},\ }\bibfield  {title} {\bibinfo {title} {Monte
  carlo simulations of two-dimensional hard core lattice gases},\ }\href@noop
  {} {\bibfield  {journal} {\bibinfo  {journal} {J. Chem. Phys.}\ }\textbf
  {\bibinfo {volume} {126}},\ \bibinfo {pages} {114508} (\bibinfo {year}
  {2007})}\BibitemShut {NoStop}%
\bibitem [{\citenamefont {Nath}\ and\ \citenamefont
  {Rajesh}(2014)}]{2014-nr-pre-multiple}%
  \BibitemOpen
  \bibfield  {author} {\bibinfo {author} {\bibfnamefont {T.}~\bibnamefont
  {Nath}}\ and\ \bibinfo {author} {\bibfnamefont {R.}~\bibnamefont {Rajesh}},\
  }\bibfield  {title} {\bibinfo {title} {Multiple phase transitions in extended
  hard-core lattice gas models in two dimensions},\ }\href
  {https://doi.org/10.1103/PhysRevE.90.012120} {\bibfield  {journal} {\bibinfo
  {journal} {Phys. Rev. E}\ }\textbf {\bibinfo {volume} {90}},\ \bibinfo
  {pages} {012120} (\bibinfo {year} {2014})}\BibitemShut {NoStop}%
\bibitem [{\citenamefont {Nath}\ and\ \citenamefont
  {Rajesh}(2016)}]{2016-nr-jsm-high}%
  \BibitemOpen
  \bibfield  {author} {\bibinfo {author} {\bibfnamefont {T.}~\bibnamefont
  {Nath}}\ and\ \bibinfo {author} {\bibfnamefont {R.}~\bibnamefont {Rajesh}},\
  }\bibfield  {title} {\bibinfo {title} {The high density phase of the k -nn
  hard core lattice gas model},\ }\href
  {http://stacks.iop.org/1742-5468/2016/i=7/a=073203} {\bibfield  {journal}
  {\bibinfo  {journal} {J. Stat. Mech.}\ }\textbf {\bibinfo {volume} {2016}},\
  \bibinfo {pages} {073203} (\bibinfo {year} {2016})}\BibitemShut {NoStop}%
\bibitem [{\citenamefont {Akimenko}\ \emph {et~al.}(2019)\citenamefont
  {Akimenko}, \citenamefont {Gorbunov}, \citenamefont {Myshlyavtsev},\ and\
  \citenamefont {Stishenko}}]{akimenko2019tensor}%
  \BibitemOpen
  \bibfield  {author} {\bibinfo {author} {\bibfnamefont {S.~S.}\ \bibnamefont
  {Akimenko}}, \bibinfo {author} {\bibfnamefont {V.~A.}\ \bibnamefont
  {Gorbunov}}, \bibinfo {author} {\bibfnamefont {A.~V.}\ \bibnamefont
  {Myshlyavtsev}},\ and\ \bibinfo {author} {\bibfnamefont {P.~V.}\ \bibnamefont
  {Stishenko}},\ }\bibfield  {title} {\bibinfo {title} {Tensor renormalization
  group study of hard-disk models on a triangular lattice},\ }\href@noop {}
  {\bibfield  {journal} {\bibinfo  {journal} {Physical Review E}\ }\textbf
  {\bibinfo {volume} {100}},\ \bibinfo {pages} {022108} (\bibinfo {year}
  {2019})}\BibitemShut {NoStop}%
\bibitem [{\citenamefont {Thewes}\ and\ \citenamefont
  {Fernandes}(2020)}]{PhysRevE.101.062138}%
  \BibitemOpen
  \bibfield  {author} {\bibinfo {author} {\bibfnamefont {F.~C.}\ \bibnamefont
  {Thewes}}\ and\ \bibinfo {author} {\bibfnamefont {H.~C.~M.}\ \bibnamefont
  {Fernandes}},\ }\bibfield  {title} {\bibinfo {title} {Phase transitions in
  hard-core lattice gases on the honeycomb lattice},\ }\href
  {https://doi.org/10.1103/PhysRevE.101.062138} {\bibfield  {journal} {\bibinfo
   {journal} {Phys. Rev. E}\ }\textbf {\bibinfo {volume} {101}},\ \bibinfo
  {pages} {062138} (\bibinfo {year} {2020})}\BibitemShut {NoStop}%
\bibitem [{\citenamefont {Vigneshwar}\ \emph {et~al.}(2019)\citenamefont
  {Vigneshwar}, \citenamefont {Mandal}, \citenamefont {Damle}, \citenamefont
  {Dhar},\ and\ \citenamefont {Rajesh}}]{vigneshwar2019phase}%
  \BibitemOpen
  \bibfield  {author} {\bibinfo {author} {\bibfnamefont {N.}~\bibnamefont
  {Vigneshwar}}, \bibinfo {author} {\bibfnamefont {D.}~\bibnamefont {Mandal}},
  \bibinfo {author} {\bibfnamefont {K.}~\bibnamefont {Damle}}, \bibinfo
  {author} {\bibfnamefont {D.}~\bibnamefont {Dhar}},\ and\ \bibinfo {author}
  {\bibfnamefont {R.}~\bibnamefont {Rajesh}},\ }\bibfield  {title} {\bibinfo
  {title} {Phase diagram of a system of hard cubes on the cubic lattice},\
  }\href@noop {} {\bibfield  {journal} {\bibinfo  {journal} {Phys. Rev. E}\
  }\textbf {\bibinfo {volume} {99}},\ \bibinfo {pages} {052129} (\bibinfo
  {year} {2019})}\BibitemShut {NoStop}%
\bibitem [{\citenamefont {Disertori}\ \emph {et~al.}(2020)\citenamefont
  {Disertori}, \citenamefont {Giuliani},\ and\ \citenamefont
  {Jauslin}}]{disertori2020plate}%
  \BibitemOpen
  \bibfield  {author} {\bibinfo {author} {\bibfnamefont {M.}~\bibnamefont
  {Disertori}}, \bibinfo {author} {\bibfnamefont {A.}~\bibnamefont
  {Giuliani}},\ and\ \bibinfo {author} {\bibfnamefont {I.}~\bibnamefont
  {Jauslin}},\ }\bibfield  {title} {\bibinfo {title} {Plate-nematic phase in
  three dimensions},\ }\href@noop {} {\bibfield  {journal} {\bibinfo  {journal}
  {Commun. Math. Phys.}\ }\textbf {\bibinfo {volume} {373}},\ \bibinfo {pages}
  {327} (\bibinfo {year} {2020})}\BibitemShut {NoStop}%
\bibitem [{\citenamefont {Mandal}\ and\ \citenamefont
  {Rajesh}(2017)}]{mandal2017columnar}%
  \BibitemOpen
  \bibfield  {author} {\bibinfo {author} {\bibfnamefont {D.}~\bibnamefont
  {Mandal}}\ and\ \bibinfo {author} {\bibfnamefont {R.}~\bibnamefont
  {Rajesh}},\ }\bibfield  {title} {\bibinfo {title} {Columnar-disorder phase
  boundary in a mixture of hard squares and dimers},\ }\href@noop {} {\bibfield
   {journal} {\bibinfo  {journal} {Phys. Rev. E}\ }\textbf {\bibinfo {volume}
  {96}},\ \bibinfo {pages} {012140} (\bibinfo {year} {2017})}\BibitemShut
  {NoStop}%
\bibitem [{\citenamefont {Kundu}\ \emph {et~al.}(2016)\citenamefont {Kundu},
  \citenamefont {Stilck},\ and\ \citenamefont {Rajesh}}]{2015-ksr-epl-phase}%
  \BibitemOpen
  \bibfield  {author} {\bibinfo {author} {\bibfnamefont {J.}~\bibnamefont
  {Kundu}}, \bibinfo {author} {\bibfnamefont {J.~F.}\ \bibnamefont {Stilck}},\
  and\ \bibinfo {author} {\bibfnamefont {R.}~\bibnamefont {Rajesh}},\
  }\bibfield  {title} {\bibinfo {title} {Phase diagram of a bidispersed
  hard-rod lattice gas in two dimensions},\ }\href
  {http://stacks.iop.org/0295-5075/112/i=6/a=66002} {\bibfield  {journal}
  {\bibinfo  {journal} {Eur. Phys. Lett.}\ }\textbf {\bibinfo {volume} {112}},\
  \bibinfo {pages} {66002} (\bibinfo {year} {2016})}\BibitemShut {NoStop}%
\bibitem [{\citenamefont {Stilck}\ and\ \citenamefont
  {Rajesh}(2015)}]{2015-sr-pre-polydispersed}%
  \BibitemOpen
  \bibfield  {author} {\bibinfo {author} {\bibfnamefont {J.~F.}\ \bibnamefont
  {Stilck}}\ and\ \bibinfo {author} {\bibfnamefont {R.}~\bibnamefont
  {Rajesh}},\ }\bibfield  {title} {\bibinfo {title} {Polydispersed rods on the
  square lattice},\ }\href {https://doi.org/10.1103/PhysRevE.91.012106}
  {\bibfield  {journal} {\bibinfo  {journal} {Phys. Rev. E}\ }\textbf {\bibinfo
  {volume} {91}},\ \bibinfo {pages} {012106} (\bibinfo {year}
  {2015})}\BibitemShut {NoStop}%
\bibitem [{\citenamefont {Rodrigues}\ and\ \citenamefont
  {Oliveira}(2019)}]{rodrigues2019three}%
  \BibitemOpen
  \bibfield  {author} {\bibinfo {author} {\bibfnamefont {N.~T.}\ \bibnamefont
  {Rodrigues}}\ and\ \bibinfo {author} {\bibfnamefont {T.~J.}\ \bibnamefont
  {Oliveira}},\ }\bibfield  {title} {\bibinfo {title} {Three stable phases and
  thermodynamic anomaly in a binary mixture of hard particles},\ }\href@noop {}
  {\bibfield  {journal} {\bibinfo  {journal} {J. Chem. Phys.}\ }\textbf
  {\bibinfo {volume} {151}},\ \bibinfo {pages} {024504} (\bibinfo {year}
  {2019})}\BibitemShut {NoStop}%
\bibitem [{\citenamefont {Zwanzig}(1963)}]{1963-z-jcp-first}%
  \BibitemOpen
  \bibfield  {author} {\bibinfo {author} {\bibfnamefont {R.}~\bibnamefont
  {Zwanzig}},\ }\bibfield  {title} {\bibinfo {title} {First‐order phase
  transition in a gas of long thin rods},\ }\href
  {https://doi.org/10.1063/1.1734518} {\bibfield  {journal} {\bibinfo
  {journal} {J. Chem. Phys.}\ }\textbf {\bibinfo {volume} {39}},\ \bibinfo
  {pages} {1714} (\bibinfo {year} {1963})},\ \Eprint
  {https://arxiv.org/abs/http://dx.doi.org/10.1063/1.1734518}
  {http://dx.doi.org/10.1063/1.1734518} \BibitemShut {NoStop}%
\bibitem [{\citenamefont {Onsager}(1949)}]{1949-o-nyas-effects}%
  \BibitemOpen
  \bibfield  {author} {\bibinfo {author} {\bibfnamefont {L.}~\bibnamefont
  {Onsager}},\ }\bibfield  {title} {\bibinfo {title} {The effects of shape on
  the interaction of colloidal particles},\ }\href
  {https://doi.org/10.1111/j.1749-6632.1949.tb27296.x} {\bibfield  {journal}
  {\bibinfo  {journal} {Ann. N. Y. Acad. Sci.}\ }\textbf {\bibinfo {volume}
  {51}},\ \bibinfo {pages} {627} (\bibinfo {year} {1949})}\BibitemShut
  {NoStop}%
\bibitem [{\citenamefont {Kundu}\ \emph {et~al.}(2012)\citenamefont {Kundu},
  \citenamefont {Rajesh}, \citenamefont {Dhar},\ and\ \citenamefont
  {Stilck}}]{2012-krds-aipcp-monte}%
  \BibitemOpen
  \bibfield  {author} {\bibinfo {author} {\bibfnamefont {J.}~\bibnamefont
  {Kundu}}, \bibinfo {author} {\bibfnamefont {R.}~\bibnamefont {Rajesh}},
  \bibinfo {author} {\bibfnamefont {D.}~\bibnamefont {Dhar}},\ and\ \bibinfo
  {author} {\bibfnamefont {J.~F.}\ \bibnamefont {Stilck}},\ }\bibfield  {title}
  {\bibinfo {title} {A monte carlo algorithm for studying phase transition in
  systems of hard rigid rods},\ }\href {https://doi.org/10.1063/1.4709907}
  {\bibfield  {journal} {\bibinfo  {journal} {AIP Conf. Proc.}\ }\textbf
  {\bibinfo {volume} {1447}},\ \bibinfo {pages} {113} (\bibinfo {year}
  {2012})}\BibitemShut {NoStop}%
\bibitem [{\citenamefont {Kundu}\ \emph {et~al.}(2013)\citenamefont {Kundu},
  \citenamefont {Rajesh}, \citenamefont {Dhar},\ and\ \citenamefont
  {Stilck}}]{2013-krds-pre-nematic}%
  \BibitemOpen
  \bibfield  {author} {\bibinfo {author} {\bibfnamefont {J.}~\bibnamefont
  {Kundu}}, \bibinfo {author} {\bibfnamefont {R.}~\bibnamefont {Rajesh}},
  \bibinfo {author} {\bibfnamefont {D.}~\bibnamefont {Dhar}},\ and\ \bibinfo
  {author} {\bibfnamefont {J.~F.}\ \bibnamefont {Stilck}},\ }\bibfield  {title}
  {\bibinfo {title} {Nematic-disordered phase transition in systems of long
  rigid rods on two-dimensional lattices},\ }\href
  {https://doi.org/10.1103/PhysRevE.87.032103} {\bibfield  {journal} {\bibinfo
  {journal} {Phys. Rev. E}\ }\textbf {\bibinfo {volume} {87}},\ \bibinfo
  {pages} {032103} (\bibinfo {year} {2013})}\BibitemShut {NoStop}%
\bibitem [{\citenamefont {Disertori}\ and\ \citenamefont
  {Giuliani}(2013)}]{2013-dg-cmp-nematic}%
  \BibitemOpen
  \bibfield  {author} {\bibinfo {author} {\bibfnamefont {M.}~\bibnamefont
  {Disertori}}\ and\ \bibinfo {author} {\bibfnamefont {A.}~\bibnamefont
  {Giuliani}},\ }\bibfield  {title} {\bibinfo {title} {The nematic phase of a
  system of long hard rods},\ }\href
  {https://doi.org/10.1007/s00220-013-1767-1} {\bibfield  {journal} {\bibinfo
  {journal} {Commun. Math. Phys.}\ }\textbf {\bibinfo {volume} {323}},\
  \bibinfo {pages} {143} (\bibinfo {year} {2013})}\BibitemShut {NoStop}%
\bibitem [{\citenamefont {Matoz-Fernandez}\ \emph
  {et~al.}(2008{\natexlab{a}})\citenamefont {Matoz-Fernandez}, \citenamefont
  {Linares},\ and\ \citenamefont
  {Ramirez-Pastor}}]{2008-mlr-epl-determination}%
  \BibitemOpen
  \bibfield  {author} {\bibinfo {author} {\bibfnamefont {D.~A.}\ \bibnamefont
  {Matoz-Fernandez}}, \bibinfo {author} {\bibfnamefont {D.~H.}\ \bibnamefont
  {Linares}},\ and\ \bibinfo {author} {\bibfnamefont {A.~J.}\ \bibnamefont
  {Ramirez-Pastor}},\ }\bibfield  {title} {\bibinfo {title} {Determination of
  the critical exponents for the isotropic-nematic phase transition in a system
  of long rods on two-dimensional lattices: Universality of the transition},\
  }\href@noop {} {\bibfield  {journal} {\bibinfo  {journal} {Eur. Phys. Lett.}\
  }\textbf {\bibinfo {volume} {82}},\ \bibinfo {pages} {50007} (\bibinfo {year}
  {2008}{\natexlab{a}})}\BibitemShut {NoStop}%
\bibitem [{\citenamefont {Matoz-Fernandez}\ \emph
  {et~al.}(2008{\natexlab{b}})\citenamefont {Matoz-Fernandez}, \citenamefont
  {Linares},\ and\ \citenamefont {Ramirez-Pastor}}]{2008-mlr-jcp-critical}%
  \BibitemOpen
  \bibfield  {author} {\bibinfo {author} {\bibfnamefont {D.~A.}\ \bibnamefont
  {Matoz-Fernandez}}, \bibinfo {author} {\bibfnamefont {D.~H.}\ \bibnamefont
  {Linares}},\ and\ \bibinfo {author} {\bibfnamefont {A.~J.}\ \bibnamefont
  {Ramirez-Pastor}},\ }\bibfield  {title} {\bibinfo {title} {Critical behavior
  of long straight rigid rods on two-dimensional lattices: Theory and monte
  carlo simulations},\ }\href {http://dx.doi.org/10.1063/1.2927877} {\bibfield
  {journal} {\bibinfo  {journal} {J. Chem. Phys.}\ }\textbf {\bibinfo {volume}
  {128}},\ \bibinfo {pages} {214902} (\bibinfo {year}
  {2008}{\natexlab{b}})}\BibitemShut {NoStop}%
\bibitem [{\citenamefont {Fischer}\ and\ \citenamefont
  {Vink}(2009)}]{2009-fv-epl-restricted}%
  \BibitemOpen
  \bibfield  {author} {\bibinfo {author} {\bibfnamefont {T.}~\bibnamefont
  {Fischer}}\ and\ \bibinfo {author} {\bibfnamefont {R.~L.~C.}\ \bibnamefont
  {Vink}},\ }\bibfield  {title} {\bibinfo {title} {Restricted orientation
  "liquid crystal" in two dimensions: Isotropic-nematic transition or
  liquid-gas one(?)},\ }\href {http://stacks.iop.org/0295-5075/85/i=5/a=56003}
  {\bibfield  {journal} {\bibinfo  {journal} {Eur. Phys. Lett.}\ }\textbf
  {\bibinfo {volume} {85}},\ \bibinfo {pages} {56003} (\bibinfo {year}
  {2009})}\BibitemShut {NoStop}%
\bibitem [{\citenamefont {Matoz-Fernandez}\ \emph
  {et~al.}(2008{\natexlab{c}})\citenamefont {Matoz-Fernandez}, \citenamefont
  {Linares},\ and\ \citenamefont {Ramirez-Pastor}}]{2008-mlr-pa-critical}%
  \BibitemOpen
  \bibfield  {author} {\bibinfo {author} {\bibfnamefont {D.}~\bibnamefont
  {Matoz-Fernandez}}, \bibinfo {author} {\bibfnamefont {D.}~\bibnamefont
  {Linares}},\ and\ \bibinfo {author} {\bibfnamefont {A.}~\bibnamefont
  {Ramirez-Pastor}},\ }\bibfield  {title} {\bibinfo {title} {Critical behavior
  of long linear k-mers on honeycomb lattices},\ }\href
  {http://www.sciencedirect.com/science/article/pii/S0378437108007127}
  {\bibfield  {journal} {\bibinfo  {journal} {Physica A}\ }\textbf {\bibinfo
  {volume} {387}},\ \bibinfo {pages} {6513 } (\bibinfo {year}
  {2008}{\natexlab{c}})}\BibitemShut {NoStop}%
\bibitem [{\citenamefont {Chatelain}\ and\ \citenamefont
  {Gendiar}(2020)}]{chatelain2020absence}%
  \BibitemOpen
  \bibfield  {author} {\bibinfo {author} {\bibfnamefont {C.}~\bibnamefont
  {Chatelain}}\ and\ \bibinfo {author} {\bibfnamefont {A.}~\bibnamefont
  {Gendiar}},\ }\bibfield  {title} {\bibinfo {title} {Absence of logarithmic
  divergence of the entanglement entropies at the phase transitions of a 2d
  classical hard rod model},\ }\href@noop {} {\bibfield  {journal} {\bibinfo
  {journal} {Eur. Phys. J. B}\ }\textbf {\bibinfo {volume} {93}},\ \bibinfo
  {pages} {134} (\bibinfo {year} {2020})}\BibitemShut {NoStop}%
\bibitem [{\citenamefont {Kundu}\ and\ \citenamefont
  {Rajesh}(2013)}]{2013-kr-pre-reentrant}%
  \BibitemOpen
  \bibfield  {author} {\bibinfo {author} {\bibfnamefont {J.}~\bibnamefont
  {Kundu}}\ and\ \bibinfo {author} {\bibfnamefont {R.}~\bibnamefont {Rajesh}},\
  }\bibfield  {title} {\bibinfo {title} {Reentrant disordered phase in a system
  of repulsive rods on a bethe-like lattice},\ }\href
  {https://doi.org/10.1103/PhysRevE.88.012134} {\bibfield  {journal} {\bibinfo
  {journal} {Phys. Rev. E}\ }\textbf {\bibinfo {volume} {88}},\ \bibinfo
  {pages} {012134} (\bibinfo {year} {2013})}\BibitemShut {NoStop}%
\bibitem [{\citenamefont {Vogel}\ \emph {et~al.}(2017)\citenamefont {Vogel},
  \citenamefont {Saravia},\ and\ \citenamefont
  {Ramirez-Pastor}}]{vogel2017phase}%
  \BibitemOpen
  \bibfield  {author} {\bibinfo {author} {\bibfnamefont {E.~E.}\ \bibnamefont
  {Vogel}}, \bibinfo {author} {\bibfnamefont {G.}~\bibnamefont {Saravia}},\
  and\ \bibinfo {author} {\bibfnamefont {A.~J.}\ \bibnamefont
  {Ramirez-Pastor}},\ }\bibfield  {title} {\bibinfo {title} {Phase transitions
  in a system of long rods on two-dimensional lattices by means of information
  theory},\ }\href@noop {} {\bibfield  {journal} {\bibinfo  {journal} {Phys.
  Rev. E}\ }\textbf {\bibinfo {volume} {96}},\ \bibinfo {pages} {062133}
  (\bibinfo {year} {2017})}\BibitemShut {NoStop}%
\bibitem [{\citenamefont {Vogel}\ \emph {et~al.}(2020)\citenamefont {Vogel},
  \citenamefont {Saravia}, \citenamefont {Ramirez-Pastor},\ and\ \citenamefont
  {Pasinetti}}]{vogel2020alternative}%
  \BibitemOpen
  \bibfield  {author} {\bibinfo {author} {\bibfnamefont {E.~E.}\ \bibnamefont
  {Vogel}}, \bibinfo {author} {\bibfnamefont {G.}~\bibnamefont {Saravia}},
  \bibinfo {author} {\bibfnamefont {A.~J.}\ \bibnamefont {Ramirez-Pastor}},\
  and\ \bibinfo {author} {\bibfnamefont {M.}~\bibnamefont {Pasinetti}},\
  }\bibfield  {title} {\bibinfo {title} {Alternative characterization of the
  nematic transition in deposition of rods on two-dimensional lattices},\
  }\href@noop {} {\bibfield  {journal} {\bibinfo  {journal} {Phys. Rev. E}\
  }\textbf {\bibinfo {volume} {101}},\ \bibinfo {pages} {022104} (\bibinfo
  {year} {2020})}\BibitemShut {NoStop}%
\bibitem [{\citenamefont {Vigneshwar}\ \emph {et~al.}(2017)\citenamefont
  {Vigneshwar}, \citenamefont {Dhar},\ and\ \citenamefont
  {Rajesh}}]{2017-vdr-jsm-different}%
  \BibitemOpen
  \bibfield  {author} {\bibinfo {author} {\bibfnamefont {N.}~\bibnamefont
  {Vigneshwar}}, \bibinfo {author} {\bibfnamefont {D.}~\bibnamefont {Dhar}},\
  and\ \bibinfo {author} {\bibfnamefont {R.}~\bibnamefont {Rajesh}},\
  }\bibfield  {title} {\bibinfo {title} {Different phases of a system of hard
  rods on three dimensional cubic lattice},\ }\href
  {http://stacks.iop.org/1742-5468/2017/i=11/a=113304} {\bibfield  {journal}
  {\bibinfo  {journal} {J. Stat. Mech.}\ }\textbf {\bibinfo {volume} {2017}},\
  \bibinfo {pages} {113304} (\bibinfo {year} {2017})}\BibitemShut {NoStop}%
\bibitem [{\citenamefont {Gschwind}\ \emph {et~al.}(2017)\citenamefont
  {Gschwind}, \citenamefont {Klopotek}, \citenamefont {Ai},\ and\ \citenamefont
  {Oettel}}]{2017-gkao-pre-isotropic}%
  \BibitemOpen
  \bibfield  {author} {\bibinfo {author} {\bibfnamefont {A.}~\bibnamefont
  {Gschwind}}, \bibinfo {author} {\bibfnamefont {M.}~\bibnamefont {Klopotek}},
  \bibinfo {author} {\bibfnamefont {Y.}~\bibnamefont {Ai}},\ and\ \bibinfo
  {author} {\bibfnamefont {M.}~\bibnamefont {Oettel}},\ }\bibfield  {title}
  {\bibinfo {title} {Isotropic-nematic transition for hard rods on a
  three-dimensional cubic lattice},\ }\href
  {https://doi.org/10.1103/PhysRevE.96.012104} {\bibfield  {journal} {\bibinfo
  {journal} {Phys. Rev. E}\ }\textbf {\bibinfo {volume} {96}},\ \bibinfo
  {pages} {012104} (\bibinfo {year} {2017})}\BibitemShut {NoStop}%
\bibitem [{\citenamefont {Wojciechowski}\ and\ \citenamefont
  {Frenkel}(2004)}]{2004-wf-cmst-tetratic}%
  \BibitemOpen
  \bibfield  {author} {\bibinfo {author} {\bibfnamefont {K.}~\bibnamefont
  {Wojciechowski}}\ and\ \bibinfo {author} {\bibfnamefont {D.}~\bibnamefont
  {Frenkel}},\ }\bibfield  {title} {\bibinfo {title} {Tetratic phase in the
  planar hard square system},\ }\href@noop {} {\bibfield  {journal} {\bibinfo
  {journal} {Comput. Meth. Sci. Technol.}\ }\textbf {\bibinfo {volume} {10}},\
  \bibinfo {pages} {235} (\bibinfo {year} {2004})}\BibitemShut {NoStop}%
\bibitem [{\citenamefont {Kasteleyn}(1961)}]{1961-k-physica-statistics}%
  \BibitemOpen
  \bibfield  {author} {\bibinfo {author} {\bibfnamefont {P.}~\bibnamefont
  {Kasteleyn}},\ }\bibfield  {title} {\bibinfo {title} {The statistics of
  dimers on a lattice},\ }\href
  {https://doi.org/http://dx.doi.org/10.1016/0031-8914(61)90063-5} {\bibfield
  {journal} {\bibinfo  {journal} {Physica}\ }\textbf {\bibinfo {volume} {27}},\
  \bibinfo {pages} {1209 } (\bibinfo {year} {1961})}\BibitemShut {NoStop}%
\bibitem [{\citenamefont {Kasteleyn}(1963)}]{kasteleyn1963dimer}%
  \BibitemOpen
  \bibfield  {author} {\bibinfo {author} {\bibfnamefont {P.~W.}\ \bibnamefont
  {Kasteleyn}},\ }\bibfield  {title} {\bibinfo {title} {Dimer statistics and
  phase transitions},\ }\href@noop {} {\bibfield  {journal} {\bibinfo
  {journal} {J. Math. Phys.}\ }\textbf {\bibinfo {volume} {4}},\ \bibinfo
  {pages} {287} (\bibinfo {year} {1963})}\BibitemShut {NoStop}%
\bibitem [{\citenamefont {Temperley}\ and\ \citenamefont
  {Fisher}(1961)}]{1961-tf-pm-dimer}%
  \BibitemOpen
  \bibfield  {author} {\bibinfo {author} {\bibfnamefont {H.~N.~V.}\
  \bibnamefont {Temperley}}\ and\ \bibinfo {author} {\bibfnamefont {M.~E.}\
  \bibnamefont {Fisher}},\ }\bibfield  {title} {\bibinfo {title} {Dimer problem
  in statistical mechanics-an exact result},\ }\href
  {http://dx.doi.org/10.1080/14786436108243366} {\bibfield  {journal} {\bibinfo
   {journal} {Phil. Mag.}\ }\textbf {\bibinfo {volume} {6}},\ \bibinfo {pages}
  {1061} (\bibinfo {year} {1961})}\BibitemShut {NoStop}%
\bibitem [{\citenamefont {Fisher}(1961)}]{1961-f-pr-statistical}%
  \BibitemOpen
  \bibfield  {author} {\bibinfo {author} {\bibfnamefont {M.~E.}\ \bibnamefont
  {Fisher}},\ }\bibfield  {title} {\bibinfo {title} {Statistical mechanics of
  dimers on a plane lattice},\ }\href
  {https://doi.org/10.1103/PhysRev.124.1664} {\bibfield  {journal} {\bibinfo
  {journal} {Phys. Rev.}\ }\textbf {\bibinfo {volume} {124}},\ \bibinfo {pages}
  {1664} (\bibinfo {year} {1961})}\BibitemShut {NoStop}%
\bibitem [{\citenamefont {Fisher}\ and\ \citenamefont
  {Stephenson}(1963)}]{1963-fs-pr-statistical}%
  \BibitemOpen
  \bibfield  {author} {\bibinfo {author} {\bibfnamefont {M.~E.}\ \bibnamefont
  {Fisher}}\ and\ \bibinfo {author} {\bibfnamefont {J.}~\bibnamefont
  {Stephenson}},\ }\bibfield  {title} {\bibinfo {title} {Statistical mechanics
  of dimers on a plane lattice. ii. dimer correlations and monomers},\ }\href
  {https://doi.org/10.1103/PhysRev.132.1411} {\bibfield  {journal} {\bibinfo
  {journal} {Phys. Rev.}\ }\textbf {\bibinfo {volume} {132}},\ \bibinfo {pages}
  {1411} (\bibinfo {year} {1963})}\BibitemShut {NoStop}%
\bibitem [{\citenamefont {Fendley}\ \emph {et~al.}(2002)\citenamefont
  {Fendley}, \citenamefont {Moessner},\ and\ \citenamefont
  {Sondhi}}]{2002-fms-prb-classical}%
  \BibitemOpen
  \bibfield  {author} {\bibinfo {author} {\bibfnamefont {P.}~\bibnamefont
  {Fendley}}, \bibinfo {author} {\bibfnamefont {R.}~\bibnamefont {Moessner}},\
  and\ \bibinfo {author} {\bibfnamefont {S.~L.}\ \bibnamefont {Sondhi}},\
  }\bibfield  {title} {\bibinfo {title} {Classical dimers on the triangular
  lattice},\ }\href {https://doi.org/10.1103/PhysRevB.66.214513} {\bibfield
  {journal} {\bibinfo  {journal} {Phys. Rev. B}\ }\textbf {\bibinfo {volume}
  {66}},\ \bibinfo {pages} {214513} (\bibinfo {year} {2002})}\BibitemShut
  {NoStop}%
\bibitem [{\citenamefont {Nagle}\ \emph {et~al.}(1989)\citenamefont {Nagle},
  \citenamefont {Yokoi},\ and\ \citenamefont {Bhattacharjee}}]{nagle1989dimer}%
  \BibitemOpen
  \bibfield  {author} {\bibinfo {author} {\bibfnamefont {J.~F.}\ \bibnamefont
  {Nagle}}, \bibinfo {author} {\bibfnamefont {C.~S.}\ \bibnamefont {Yokoi}},\
  and\ \bibinfo {author} {\bibfnamefont {S.~M.}\ \bibnamefont
  {Bhattacharjee}},\ }\bibfield  {title} {\bibinfo {title} {Dimer models on
  anisotropic lattices},\ }\href@noop {} {\bibfield  {journal} {\bibinfo
  {journal} {Phase transitions and critical phenomena}\ }\textbf {\bibinfo
  {volume} {13}},\ \bibinfo {pages} {235} (\bibinfo {year} {1989})}\BibitemShut
  {NoStop}%
\bibitem [{\citenamefont {Dhar}\ and\ \citenamefont
  {Chandra}(2008)}]{2008-dc-prl-exact}%
  \BibitemOpen
  \bibfield  {author} {\bibinfo {author} {\bibfnamefont {D.}~\bibnamefont
  {Dhar}}\ and\ \bibinfo {author} {\bibfnamefont {S.}~\bibnamefont {Chandra}},\
  }\bibfield  {title} {\bibinfo {title} {Exact entropy of dimer coverings for a
  class of lattices in three or more dimensions},\ }\href
  {https://doi.org/10.1103/PhysRevLett.100.120602} {\bibfield  {journal}
  {\bibinfo  {journal} {Phys. Rev. Lett.}\ }\textbf {\bibinfo {volume} {100}},\
  \bibinfo {pages} {120602} (\bibinfo {year} {2008})}\BibitemShut {NoStop}%
\bibitem [{\citenamefont {Ghosh}\ \emph {et~al.}(2007)\citenamefont {Ghosh},
  \citenamefont {Dhar},\ and\ \citenamefont {Jacobsen}}]{2007-gdj-pre-random}%
  \BibitemOpen
  \bibfield  {author} {\bibinfo {author} {\bibfnamefont {A.}~\bibnamefont
  {Ghosh}}, \bibinfo {author} {\bibfnamefont {D.}~\bibnamefont {Dhar}},\ and\
  \bibinfo {author} {\bibfnamefont {J.~L.}\ \bibnamefont {Jacobsen}},\
  }\bibfield  {title} {\bibinfo {title} {Random trimer tilings},\ }\href
  {https://doi.org/10.1103/PhysRevE.75.011115} {\bibfield  {journal} {\bibinfo
  {journal} {Phys. Rev. E}\ }\textbf {\bibinfo {volume} {75}},\ \bibinfo
  {pages} {011115} (\bibinfo {year} {2007})}\BibitemShut {NoStop}%
\bibitem [{\citenamefont {Kenyon}(2000)}]{kenyon2000conformal}%
  \BibitemOpen
  \bibfield  {author} {\bibinfo {author} {\bibfnamefont {R.}~\bibnamefont
  {Kenyon}},\ }\bibfield  {title} {\bibinfo {title} {Conformal invariance of
  domino tiling},\ }\href@noop {} {\bibfield  {journal} {\bibinfo  {journal}
  {Ann. Prob.}\ ,\ \bibinfo {pages} {759}} (\bibinfo {year}
  {2000})}\BibitemShut {NoStop}%
\bibitem [{\citenamefont {Gagunashvili}\ and\ \citenamefont
  {Priezzhev}(1979)}]{gagunashvili1979close}%
  \BibitemOpen
  \bibfield  {author} {\bibinfo {author} {\bibfnamefont {N.~D.}\ \bibnamefont
  {Gagunashvili}}\ and\ \bibinfo {author} {\bibfnamefont {V.~B.}\ \bibnamefont
  {Priezzhev}},\ }\bibfield  {title} {\bibinfo {title} {Close packing of
  rectilinear polymers on a square lattice},\ }\href
  {https://doi.org/https://doi.org/10.1007/BF01017997} {\bibfield  {journal}
  {\bibinfo  {journal} {Theor Math Phys}\ }\textbf {\bibinfo {volume} {39}},\
  \bibinfo {pages} {507–510} (\bibinfo {year} {1979})}\BibitemShut {NoStop}%
\bibitem [{\citenamefont {Corless}\ \emph {et~al.}(1996)\citenamefont
  {Corless}, \citenamefont {Gonnet}, \citenamefont {Hare}, \citenamefont
  {Jeffrey},\ and\ \citenamefont {Knuth}}]{corless1996lambertw}%
  \BibitemOpen
  \bibfield  {author} {\bibinfo {author} {\bibfnamefont {R.~M.}\ \bibnamefont
  {Corless}}, \bibinfo {author} {\bibfnamefont {G.~H.}\ \bibnamefont {Gonnet}},
  \bibinfo {author} {\bibfnamefont {D.~E.}\ \bibnamefont {Hare}}, \bibinfo
  {author} {\bibfnamefont {D.~J.}\ \bibnamefont {Jeffrey}},\ and\ \bibinfo
  {author} {\bibfnamefont {D.~E.}\ \bibnamefont {Knuth}},\ }\bibfield  {title}
  {\bibinfo {title} {On the lambert-w function},\ }\href@noop {} {\bibfield
  {journal} {\bibinfo  {journal} {Advances in Computational mathematics}\
  }\textbf {\bibinfo {volume} {5}},\ \bibinfo {pages} {329} (\bibinfo {year}
  {1996})}\BibitemShut {NoStop}%
\bibitem [{\citenamefont {Nath}\ \emph {et~al.}(2015)\citenamefont {Nath},
  \citenamefont {Kundu},\ and\ \citenamefont {Rajesh}}]{2015-nkr-jsp-high}%
  \BibitemOpen
  \bibfield  {author} {\bibinfo {author} {\bibfnamefont {T.}~\bibnamefont
  {Nath}}, \bibinfo {author} {\bibfnamefont {J.}~\bibnamefont {Kundu}},\ and\
  \bibinfo {author} {\bibfnamefont {R.}~\bibnamefont {Rajesh}},\ }\bibfield
  {title} {\bibinfo {title} {High-activity expansion for the columnar phase of
  the hard rectangle gas},\ }\href {https://doi.org/10.1007/s10955-015-1285-y}
  {\bibfield  {journal} {\bibinfo  {journal} {J. Stat. Phys.}\ }\textbf
  {\bibinfo {volume} {160}},\ \bibinfo {pages} {1173} (\bibinfo {year}
  {2015})}\BibitemShut {NoStop}%
\end{thebibliography}


\end{document}